\documentclass[twocolumn,superscriptaddress,floatfix,prx]{revtex4}
\usepackage{graphicx,amsfonts,amssymb,amsmath,hyperref,hypcap,enumerate}
\usepackage{slashed}
\usepackage{tikz}
\usepackage{amsthm}

\newcommand{\be}{\begin{equation}}
\newcommand{\ee}{\end{equation}}

\hypersetup{%
pdftitle={Locality of Greens},%
pdfauthor={Inti Sodemann},%
pdfpagemode={UseNone},%
pdfstartview={FitH},%
breaklinks=true}

\begin{document}
\title{Quantum Hall ferroelectrics and nematics in multivalley systems}

\author{Inti Sodemann}
\affiliation{Department of Physics, Massachusetts Institute of Technology, Cambridge, Massachusetts 02139, USA}
\affiliation{Max-Planck Institute for the Physics of Complex Systems, D-01187 Dresden, Germany}
\author{Zheng Zhu}
\affiliation{Department of Physics, Massachusetts Institute of Technology, Cambridge, Massachusetts 02139, USA}
\author{Liang Fu}
\affiliation{Department of Physics, Massachusetts Institute of Technology, Cambridge, Massachusetts 02139, USA}

\date{\today}

\begin{abstract}
We study broken symmetry states at integer Landau level fillings in multivalley quantum Hall systems whose low energy dispersions are anisotropic. When the Fermi surface of individual pockets lacks twofold rotational symmetry, like in Bismuth (111)~\cite{Yazdani} and in Sn$_{1-x}$Pb$_x$Se (001)~\cite{PbSnSe} surfaces, interactions tend to drive the formation of quantum Hall ferroelectric states. We demonstrate that the dipole moment in these states has an intimate relation to the Fermi surface geometry of the parent metal. In quantum Hall nematic states, like those arising in AlAs quantum wells, we demonstrate the existence of unusually robust skyrmion quasiparticles.
\end{abstract}
\maketitle

\section{Introduction}

Quantum Hall liquids with nearly degenerate internal degrees of freedom have long been the source of a rich variety of phenomena. Aside from  multilayer systems, there exist today an array of multi-valley two-dimensional electron systems exhibiting quantum Hall effect, including quantum wells of AlAs~\cite{AlAsrev}, monolayer and bilayer graphene~\cite{Kim,Novoselov}, Silicon surfaces~\cite{Si111}, PbTe (111) surfaces~\cite{PbTe}, surface states of the topological crystalline insulator Sn$_{1-x}$Pb$_x$Se~\cite{Madhavan}, and more recently the (111) surface of Bismuth~\cite{Yazdani}. These systems all have multiple valleys related to each other by discrete crystal symmetries.


%

Among these multivalley systems,  AlAs heterostructures, Si surfaces, PbTe (111) quantum wells and Bi(111) surfaces all have highly anisotropic pockets, whose orientations are valley-dependent. The shape of the Fermi surface is crucial in determining the pattern of valley symmetry breaking. For AlAs, and PbTe, the pockets are centered at time-reversal-invariant momenta, and therefore posses an elliptical shape with two-fold symmetry. This Fermi surface anisotropy favors valley-polarized states where a subset of valley degenerate Landau levels are fully occupied ~\cite{Abanin,AllanSnTe}. This state spontaneously breaks the larger crystal rotational symmetry down to twofold rotations, and therefore the resulting state is a nematic quantum Hall state.

However, the recently studied Bi(111) surface and Sn$_{1-x}$Pb$_x$Se (001) surface, brings in a novel ingredient into this problem. As shown in (Fig.~\ref{BiARPES}), Bi (111) surface has six tadpole-shaped hole pockets~\cite{Bi2001,Bi111}; Sn$_{1-x}$Pb$_x$Se (001) surface has four crescent-shaped pockets~\cite{PbSnSe}. These pockets come in time-reversed pairs located at opposite momenta away from time-reversal-invariant points in the Brillouin zone. Importantly, the Fermi surface of each pocket does {\it not} have a twofold rotational symmetry.
Therefore, we expect Landau orbitals associated with each valley generically to carry an in-plane dipolar component.
This implies that a valley-polarized state will be an insulator that spontaneously breaks inversion symmetry. We will refer to this phase as a quantum Hall {\it ferroelectric} state. The giant fermi surface anisotropy of Bi(111) and Sn$_{1-x}$Pb$_x$Se (001) surface states makes them promising candidate systems to study this novel ferroelectricity in quantum Hall states (Fig.~\ref{BiARPES}).

In this letter we provide a unified description of these ferroelectric and nematic states in multi-valley integer quantum Hall systems. We establish that due to the Fermi surface anisotropy, the long-range Coulomb interaction generically favors states with full valley polarization.
We compute the electric polarization of quantum Hall ferroelectrics using a quantum mechanical approach based on Berry phase and also using a semiclassical approach that directly relates the dipole moment to the underlying Fermi surface geometry. For quantum Hall nematic states, we study the energetics of Skyrmion-type charged excitations~\cite{Sondhi} using the density-matrix renormalization group method, and find they are suprisingly robust against mass anisotropy of the valleys, shedding new insights on the experiments on AlAs quantum wells~\cite{AlAsskyrm}.


\section{General setup}

We start from symmetry considerations on two-dimensional (2D) multivalley systems. We consider systems with three, four, or six fold rotational symmetry, which rules out any principal axis within the 2D plane. In order for each valley to have an anisotropic dispersion, we require that the symmetry group that leaves each valley invariant, or the ``little group", is only a subgroup of $C_{2v}$, or equivalently, contains no more than a twofold rotation $(x,y)\rightarrow (-x,-y)$ and at most two mirror planes that are orthogonal to each other, $(x,y)\rightarrow(-x,y)$ and $(x,y)\rightarrow (x,-y)$.


Within this wide class of systems it is convenient to distinguish two types two-dimensional systems with multiple anisotropic valleys: type-I are those whose little group contains at most a single mirror plane; type-II are those with larger symmetry. For systems of type-I, each valley is of such low symmetry that the electron dispersion at zero field has no inversion center, i.e., $\epsilon({\bf k}) \neq \epsilon(-{\bf k})$ where ${\bf k}$ is the ``small'' momentum within the valley.
As we will show, in the quantum Hall regime and at odd-integer fillings, Coulomb interaction tend to induce a valley-polarized state 
which breaks all rotational symmetry of the crystal and is therefore a ferroelectric. For systems of the type-II, the electron dispersions at each valley have twofold symmetry $\epsilon({\bf k}) = \epsilon(-{\bf k})$, and the resulting valley polarized state will retain $\pi$ rotations. In this case,
valley-polarized quantum Hall states exhibit nematic instead of ferroelectric orders.
The Bi(111) surface~\cite{Yazdani} is a representative of type-I systems and the AlAs quantum well~\cite{AlAsrev} a representative of type-II.

To study interacting electrons in multi-valley quantum Hall systems we assume that the magnetic field is large enough so that the Hamiltonian can be projected into a set of $M$ degenerate Landau levels associated with different valleys. The system is at some integer filling $\nu\in \mathbb{Z}$ ($\nu \leq M$). The interaction between an electron with position ${\bf r}_1$ from valley $i$ and an electron with position ${\bf r}_2$ from valley $j$ is dominated by the {\it long-range} part of the Coulomb interaction~\footnote{We neglect short distance valley dependent corrections. Note also that $v$ is independent of the relative orientation of the two electrons, because systems considered here all have isotropic dielectric properties.}:

\be
v({\bf r}_1-{\bf r}_2)=\frac{e^2}{\epsilon|{\bf r}_1-{\bf r}_2|}.
\ee

\noindent Despite the bare Coulomb interaction being valley independent, it effectively becomes valley dependent after projection into the Landau levels of interest:

\be\label{P0v}
P_{LL}v({\bf r}_1-{\bf r}_2) P_{LL}=\frac{1}{A} \sum_{\bf q} v_{\bf q} F_I({\bf q}) F_J(-{\bf q}) e^{i {\bf q}\cdot({\bf R}_1-{\bf R}_2)},
\ee

\noindent where ${\bf R} \equiv {\bf r}-l^2 \hat{\bf z} \times {\bf p}$ is the intra-Landau level guiding center operator, ${\bf p}={\bf \nabla}/i-e {\bf A}$ is the mechanical momentum, and $F_J({\bf q})\equiv \langle J | e^{-i l^2  \hat{\bf z}\cdot {\bf q} \times {\bf p}} |J \rangle$ is the form factor determined by the wavefunction, $|J \rangle$, of the Landau level of interest associated with valley $J$ ($\hbar\equiv 1$, $l^2\equiv1/eB$). These form factors are crucial for the energetics of valley symmetry breaking quantum Hall states and are described for specific cases of type-I and type-II multi-valley systems in Appendices~\ref{parabolic},~\ref{cones} and~\ref{tilted}.

We would like to note that our approach rests on the assumption that the long range Coulomb force projected into a set of valley-degenerate Landau levels is the dominant term that is able to break the many-body degeneracy and select the ground states. This assumption is justified in systems with anisotropic Fermi surfaces (which are the subject of this work), provided that short-range interactions are small. For systems with nearly isotropic Fermi surfaces such as graphene, long range Coulomb interaction leaves large remnant continuous symmetries, hence short distance corrections are needed to select the ground state (see e.g. Refs.~\onlinecite{Kharitonov,Murthy}). Additionally, for massless Dirac fermions systems the Landau level projection is only perturbatively enforced by the smallness of the effective fine structure constant and hence corrections beyond naive degenerate perturbation theory that include the role of the negative energy sea might be important when such parameter is not small.

\section{Hartree-Fock Theory}

We study the Hamiltonian (\ref{P0v}) using both the Hartree-Fock approximation and the DMRG numerical method. Within Hartree-Fock, we consider translationally invariant trial Slater determinant states that are arbitrary coherent superpositions of the $M$ valleys. These states can be parametrized as follows:

\be\label{Slater}
|\Psi\rangle=\prod_{a=1}^\nu \prod_{k=1}^{N_\phi} \left( \sum_{I=1}^M \langle \chi_a |I\rangle c^\dagger_{k I} \right) |{\rm O}\rangle,
\ee

\noindent where $|{\rm O}\rangle$ is the reference vacuum in which the Landau levels of interest are empty, $k$ is an intra Landau level guiding center label, $I$ labels the $M$ valleys,  $| \chi_a \rangle$ are $N$ orthonormal vectors describing the occupied coherent combinations of such valleys. The expectation value of the energy (with the Hartree part removed by including a neutralizing background) is:



\begin{equation}\label{xchange}
\begin{split}
& E[P]=-N_\phi \sum_{I,J} X_{IJ} \langle I | P | J \rangle \langle J | P | I \rangle, \\
& P= \sum_{a=1}^\nu | \chi_a \rangle \langle \chi_a |, \ X_{IJ}=\frac{1}{2}\int\frac{d^2 {\bf q}}{(2\pi)^2} v_{\bf q} F_I({\bf q}) F_J^*({\bf q}),
\end{split}
\end{equation}

\noindent where $I$ labels the valleys and  $| \chi_a \rangle$ are $\nu$ orthonormal vectors describing the occupied coherent combinations of the valleys. Trial states are found by minimizing $E[P]$ as a function of $P$. 

\begin{figure}
	\begin{center}
		\includegraphics[width=3.3in]{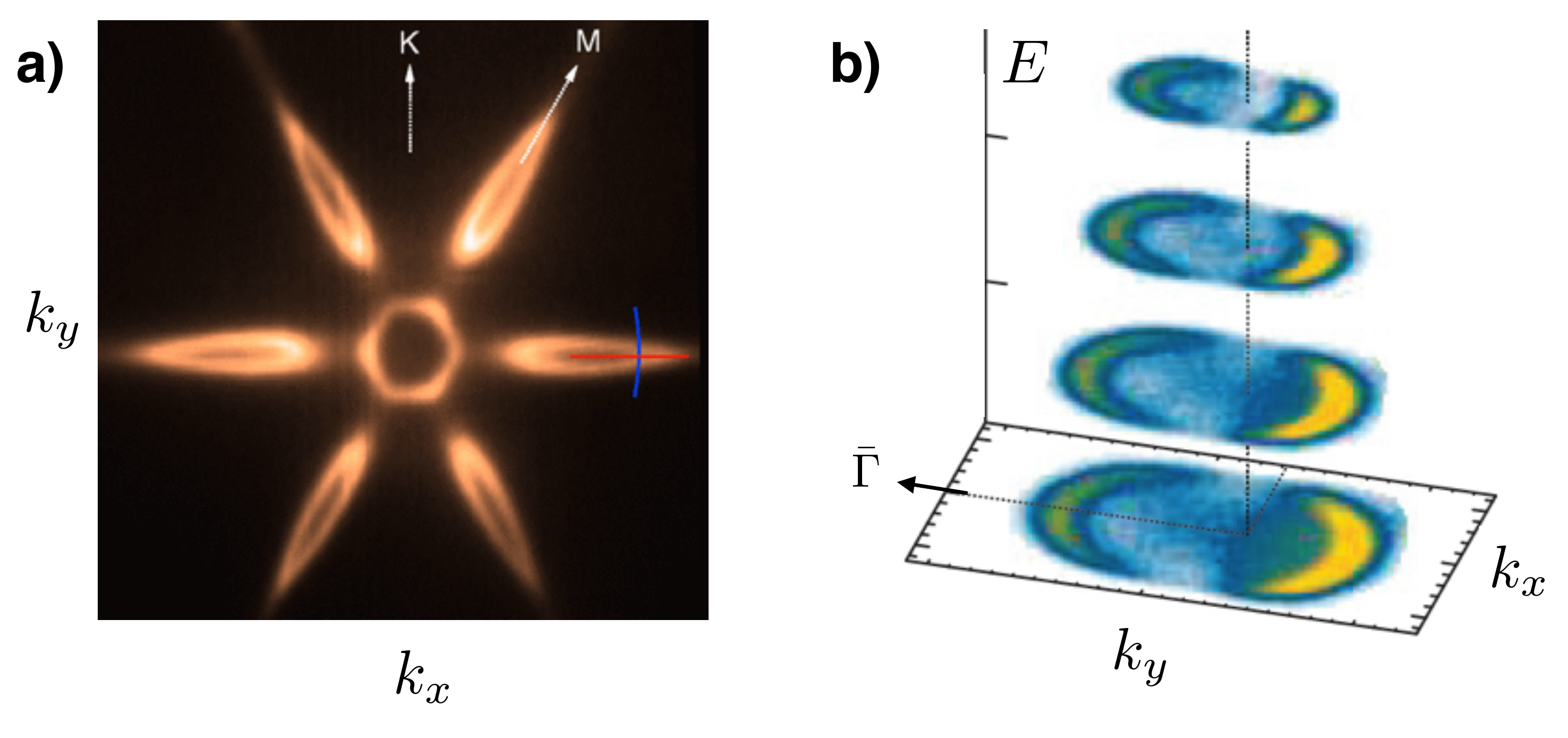}
	\end{center}
	\caption{(color online) ARPES fermi surfaces of a) Bi(111) surface from Ref~\onlinecite{Bi111} and b) Sn$_{1-x}$Pb$_x$Se (001) surface from Ref~\onlinecite{PbSnSe}.}
	\label{BiARPES}
\end{figure}

It is well known that generically at integer fillings the ground states of multicomponent systems with repulsive interactions are spontaneous integer quantum Hall states with coherence between different components known as quantum Hall ferromagnets. We have found a generalized and stronger version of quantum Hall ferromagnetism at $\nu=1$ for the kind of multicomponent systems with anisotropic fermi surfaces that we are considering. In particular, we will show that the state that minimizes the Hartree-Fock energy is maximally polarized into a single valley and those states with coherence between different valleys are energetically disfavored by the exchange energy. In fact, we will prove the following {\it theorem}:

{\color{blue} {\it Full valley polarization theorem}}--- If the set of form factors $F_I(\bf q)$, $I=\{1,..., \nu\}$, are linearly independent functions of ${\bf q}$ and the interaction is strictly repulsive, $v_{\bf q}>0$ for all ${\bf q}$, the minimum of Hartree-Fock energy $E[P]$ at $\nu=1$ for any multi-valley system is a state maximally polarized into a single valley.

{\color{blue} {\it Proof}}---We begin by noting that provided the interactions are repulsive, $v_{\bf q}>0$, the exchange integral defines an inner product for form factors as functions of ${\bf q}$:

\be
\langle F_I , F_J \rangle \equiv X_{IJ}=\frac{1}{2}\int\frac{d^2 {\bf q}}{(2\pi)^2} v_{\bf q} F_I({\bf q}) F_J^*({\bf q}).
\ee

\noindent $X_{IJ}$ is a real symmetric matrix whose entries are the inner products, known in linear algebra as a {\it Gram matrix}. A classic theorem establishes that if the vectors used in the Gram matrix are linearly independent, the matrix is strictly positive definite, namely, all its eigenvalues are real and positive. Therefore, we conclude that the linear independence of $F_I({\bf q})$ implies that $X_{IJ}$ is a strictly positive matrix.

\begin{figure}
	\begin{center}
		\includegraphics[width=2.3in]{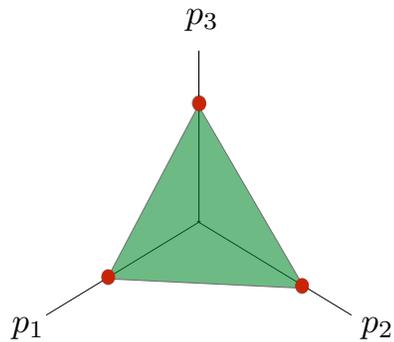}
	\end{center}
	\caption{(color online) The green triangular region defines the allowed values of $p_I$ and the red dots at the corners represent the fully valley polarized states in the case of $M=3$.}
	\label{simplex}
\end{figure}

On the other hand, at $\nu=1$, the expression for the exchange energy reduces to:

\be
E=-N_\phi \sum_{IJ} p_I X_{IJ}  p_J, \ p_I= |\langle I | \chi \rangle|^2,
\ee

\noindent where $| \chi \rangle$ is the single trial coherent combination of the valleys that is occupied at $\nu=1$. The task is to find the minimum of this energy as a function of probabilities $p_I$ that define a point in $\mathbb{R}^M$ that is further restricted to a region defined by the constraints $\sum_I p_I=1$, $p_I \in [0,1]$. Let us denote this region by $\mathbf{P}$, which is illustrated in Fig.~\ref{simplex}. Now, since $X_{IJ}$ is strictly positive definite, it defines itself an inner product in ${\mathbb R}^M$:

\be
{\bf p}_1\cdot {\bf p}_2\equiv \sum_{IJ} p_{1I} X_{IJ} p_{2J}.
\ee

\noindent Now consider a line in ${\mathbb R}^M$ parametrized by $t\in[0,1]$, ${\bf p}(t)=t {\bf p}_1+(1-t) {\bf p}_2$, connecting two distinct points, ${\bf p}_1\neq {\bf p}_2$. It is easy to show that if ${\bf p}_1\in {\bf P}$ and ${\bf p}_2\in {\bf P}$, then ${\bf p}(t)\in {\bf P}$,  $\forall t\in[0,1]$. Notice that:

\begin{equation}\label{xx}
\begin{split}
||{\bf p}(t)||^2& \equiv  {\bf p}(t)\cdot {\bf p}(t)=\\
&||{\bf p}_1-{\bf p}_2||^2 t^2 + 2 t ({\bf p}_1\cdot {\bf p}_2 -||{\bf p}_2||^2)+||{\bf p}_2||^2.\\
\end{split}
\end{equation}

\noindent Since the coefficient of $t^2$ is strictly positive, then, the maximum of $||{\bf p}(t)||^2$ is always achieved at the end points of the line and {\it never strictly within its interior}.  Since the energy is $E=-N_\phi ||{\bf p}||^2$, it follows that any point that can be viewed as being {\it strictly in the interior} of a line contained in ${\bf P}$ cannot be a minimum of the energy. Therefore the minima of the energy must necessarily be located at the corners of $\mathbf{P}$, because these are the only points which cannot be viewed as being {\it strictly in the interior} of a line contained within ${\bf P}$. The  $J$-th corner is defined by $p_J=1$ and $p_I=0$ for $I\neq J$. The state with the minimum energy is therefore fully valley polarized and the valley with minimum energy will be the one with the largest exchange integral $X_{JJ}$. In the case in which the valleys are related by a discrete crystal symmetry the exchange integral $X_{JJ}$ will be the same for every $J$ and we get a discrete degeneracy of all the fully polarized states. And thus we conclude our proof that the system spontaneously fully polarizes into a single valley.


A simple way to test that form factors are linearly independent is to check that $\det(X_{IJ})\neq0$. It is easy to imagine that form factors that arise from valleys with anisotropic Fermi surfaces are generically linearly independent. This behavior is the generic case for valleys with anisotropic Fermi surfaces that are {\it distinct}. We use the word {\it distinct} here in the strong sense that the fermi surfaces are not identical upon translation in momentum space. For example, two elliptical Fermi surfaces that are rotated with respect to each other, as in AlAs, are {\it distinct}. Similarly, all the Fermi surfaces at the Bi(111) surface are {\it distinct} (Fig.~\ref{BiARPES}). Therefore, we expect generically that the ground states of type-I and type-II systems are fully valley polarized ferroelectric or nematic states at $\nu=1$.

It is possible, however, that two or more Fermi surfaces for distinct valleys are, in a small momentum approximation, identical (as in the case of graphene within the leading linear $k$ ``dot" $p$ Dirac approximation), hence the form factors are linearly dependent. These system in the presence of long-range Coulomb interaction have an emergent SU($N$) symmetry, between the $N$ valleys with identical form factors.
Also, an important condition for the preceding analysis to apply is that all electrons reside in the same layer so that the interactions before projecting to the Landau level are identical. Otherwise, non-trivial Hartree terms may favor superpositions of components in different layers, as it happens for the exciton condensate in quantum Hall bilayers~\cite{Moon}.


In fact, the fully valley polarized states are exact eigenstates of the many-body Hamiltonian projected into the Landau levels of interest. This follows from the constraint of valley number conservation in our model implies that, after projection into the multivalley Landau level of interest, they have no other states in the Hilbert space to mix with. It is more subtle to demonstrate that they are the lowest energy states. As we have seen, however, this is the expectation based on Hartree-Fock theory. We have confirmed via density matrix renormalization group numerical simulations on the torus geometry~\cite{Shibata2001,Feiguin2008,Xiang1996,White1992,Zhao2011}, that they are indeed the exact ground states, as discussed in Section~\ref{nemat}.

Although the valley polarization theorem predicts fully valley polarized states in a general class of systems one should be mindful of potential subtleties in specific systems that would require going beyond the simple model we have described. One consideration is the importance of short-distance corrections to the Coulomb interaction, which typically are reduced from the energy scale of the Coulomb interaction by a factor of the order of $a/l$, where $a$ is a lattice length scale. A simple criterion for when these terms are important is to compare their size with the energy difference per particle between a valley polarized state and valley coherent state within the description we have provided. When such difference is small the tendency to select a unique ground state by the Coulomb interaction alone would be weak. In such cases one should consider the role of these short interactions in selecting ground states on an equal footing to the asymmetry of the form factors we have described, in an analogous fashion to how it has been done in Graphene~\cite{Kharitonov}.




\section{Quantum Hall ferroelectric dipole moment}\label{moment}

Following our full valley polarization theorem, we expect Type-I systems to exhibit valley polarized states at filling $\nu=1$ that fully breaks rotational symmetry. The presence of a macroscopic dipole moment is a subtle issue. Unlike, conventional ferroelectrics, quantum Hall ferroelectrics are always accompanied by a chiral metallic edge. Therefore one expects that any polarization charge that accumulates at the boundary of the sample to be screened by the metallic edge~\footnote{We thank A.H. MacDonald for pointing out to us this subtlety on the measurement of the Dipole moment for this state.}. In the forthcoming discussion we will ignore this subtlety in the measurement of the dipole moment. We will discuss in Section~\ref{meas} how the dipole moment that we will compute can be experimentally measured from the charge distribution of quasi-particles.

\subsection{General expressions for the dipole moment}\label{expdipole}

The dipole of the insulating state at $\nu=1$ can be defined as the change of the average position of the electrons relative to a reference state with inversion symmetry. Since the state at $\nu=1$ is a Slater determinant the dipole moment per electron can be computed for each single orbital as a single particle property. We will illustrate two different ways of computing the dipole moment which result in identical final results but are useful as they illustrate different points of view on the quantum mechanical theory of polarization.

The first approach starts from the expression that the dipole moment associated with an orbital $k$ in a valley $J$ is given by the change of the expectation value of the position operator:

\be
{\bf D}= -|e| (\langle k,J| {\bf r} | k,J \rangle-\langle k,J| {\bf r} | k,J \rangle_0).
\ee

\noindent Here $k$ is an intra-Landau level guiding center label and $J$ labels the valley that is spontaneously chosen in the ground state. For purposes of defining the dipole moment we consider a hypothetical reference hamiltonian that has inversion symmetry, whose eigenfunctions are $| k,J \rangle_0$, and imagine that this Hamiltonian can be adiabatically deformed into the Hamiltonian of interest describing the dispersion of valley $J$ which breaks inversion symmetry. The dipole is obtained as the change in the position relative to the reference inversion symmetric state. The position operator can be decomposed into cyclotron and guiding center components as follows:

\be\label{R}
{\bf r} \equiv {\bf R}- \hat{{\bf z}} \times {\bf p}.
\ee

\noindent where ${\bf p}=\nabla /i -{\bf A}({\bf r})$ is the gauge invariant mechanical momentum. The key observation is that the guiding center variables are {\it adiabatic invariants} because guiding center operators commute with the mechanical momentum, $[R_a,p_b]=0$, and the Hamiltonian describing each valley is a function of ${\bf p}$ throughout the entire adiabatic path because the path is assumed not to break translational symmetry. Thus, by choosing the origin of coordinates to be the initial position of the orbital, so that $\langle k,J| {\bf r} | k,J \rangle_0=0$, we arrive at the following expression for the dipole moment:

\be\label{Deq1}
{\bf D}= -|e| l^2  \hat{{\bf z}} \times  \langle J| {\bf p} | J \rangle.
\ee

\noindent Here we have omitted explicit reference to the guiding center label of the eigenstate in question as the formula makes manifest that the dipole moment is independent from it. The formula just derived offers great versatility for computations of the dipole moment for specific systems when the cyclotron eigenstates are known explicitly in terms of the canonical cyclotron raising and lowering operators, because ${\bf p}$ is a simple linear function of these operators. We will take advantage of this to perform a swift computation of the dipole moment in tilted Dirac cones that are relevant for the surface of topological crystalline insulators described in Section~\ref{TCI}.

The second approach follows from adapting the formalism of the modern theory of polarization~\cite{Thouless,Vanderbilt} to our current problem. To compute the dipole moment along a given direction within this approach, we chose a gauge so that there is translational invariance along such direction. For example, to compute the dipole along $y$, we choose $A_x=0$ and $A_y=Bx$. In this case the single particle states within the Landau level of interest can be labeled by a momentum $k_y$ and have a real space form:

\be
\psi_{k_y}(x,y)=\frac{e^{i k_y y}}{\sqrt{L_y}} u_{k_y}(x).
\ee

\noindent Periodicity along the $y$ direction imposes a discretization of $k_y$ ($\Delta k_y=2\pi /L_y$), and the finite size and the periodicity of the system along the $x$ direction determines the size of the effective the Brillouin zone to be: $k_y\in [0,L_x/l^2]$. To determine the polarization we compute the integral over time of the current operator:

\be
I_y=-|e| \frac{\partial H_{k_y}}{\partial k_y},
\ee

\noindent in response to an adiabatically changing Dirac cone tilt $\delta v_x(t)$ over a period of time $T$, so that $\delta v_x(t=0)=0$ and $\delta v_x (t=T)=\delta v_x$. Following the arguments of Refs.~\onlinecite{Thouless,Vanderbilt}, one finds the following expression for the change in the dipole moment per particle:

\be
D_y=-i |e| \frac{l^2}{L_x} \int_0^{L_x/l^2} dk_y \langle u_{k_y} |\partial_{k_y} |u_{k_y}\rangle.
\ee

\noindent In the present case, the real space wave-functions at different $k_y$ differ by a translation: $u_{k_y}(x)=u_0(x-k_y l^2)$. As a consequence the formula reduces to:

\begin{equation}\label{Pol}
\begin{split}
D_y&= -i |e| \langle u_{k_y} |\partial_{k_y} |u_{k_y}\rangle |_{k_y=0} \\
&=i|e| l^{2} \int d x \ u_0^*(x) \partial_x u_0(x).\\
\end{split}
\end{equation}

\noindent To make manifest the equivalence of the two approaches to compute the dipole moment, we recast Eq.~\eqref{Pol} as follows:

\begin{equation}
\begin{split}
D_y&= i|e| l^{2} \int d^2 r \ \psi^*_{k_y}(x,y) \partial_x \psi_{k_y}(x,y)=|e| l^2 \langle \psi_{k_y}| p_x |\psi_{k_y}\rangle\\
&=-|e| \langle \psi_{k_y}| (y-R_y)  |\psi_{k_y}\rangle.\\
\end{split}
\end{equation}

%

\noindent The last expression makes manifest that the dipole moment computed in this way coincides with the average position of the single particle orbital measured with respect to the guiding center $R_y$.

\subsection{WKB approximation for the dipole moment}

We now introduce a semiclassical approach to establish a direct connection of the dipole moment with the underlying Fermi surface geometry at zero field. Consider a valley described by a single-band Hamiltonian $H(p_x,p_y)$ of arbitrary form. For simplicity we will imagine that the eigenstates near this valley have negligible Berry phase, or equivalently that the Hamiltonian has only differential operators but no pseudo-spin matrix structure. This is typically a good approximation near the bottom of a band for a single-band system provided other bands are sufficiently far in energy. As in the previous subsection, by choose the Landau gauge  $A_x=0$ and $A_y=Bx$, we view the eigenvalue problem as effectively one-dimensional. The eigenvalue problem reads as:

\be\label{Hwkb}
H\left(\frac{\hbar}{i} \frac{d}{dx},k_y-eBx\right) u_{k_y}(x)=\epsilon_n u_{k_y}(x).
\ee

\noindent without loss of generality we can set $k_y=0$, since other solutions are obtained by a global translation. We can search for an approximate WKB solution to this differential equation which formally is an expansion in $\hbar$:

\be
u_0(x)= \exp \left( \frac{i}{\hbar} \phi_0(x)+\phi_1(x)+\mathcal{O}(\hbar) \right),
\ee

\noindent substituting this expression into Eq.~\eqref{Hwkb} and keeping corrections up to linear order in $\hbar$ leads to the WKB approximation of the problem~\cite{Falkovs}:

\be
u_0(x)\approx \sum_{s} \frac{1}{\sqrt{v_{x,s}^{\rm cl}(x,\epsilon)}} \exp \left( \frac{i}{\hbar} \int^x dx' p^{\rm cl}_{x,s}(x',\epsilon) \right),
\ee

\noindent where $p^{\rm cl}_{x,s}(x',\epsilon)$ is the $s$ root of the algebraic equation defined by the classical dispersion relation: $H(p^{\rm cl}_{x,s},-eBx)=\epsilon$, and

\be
v_{x,s}^{\rm cl}(x,\epsilon)=\frac{\partial H(p_x,-eBx)}{\partial p_x}|_{p_x\rightarrow p^{\rm cl}_{x,s}(x,\epsilon)},
\ee

\noindent is the group velocity evaluated on the classical trajectory. Typically we have two roots $s={+,-}$ and two turning points $x_{+,-}$ that separate the classically allowed from the classically forbidden regions. In the classically forbidden regions $p^{\rm cl}_{x,s}$ is imaginary and the wavefunction decays exponentially. As is well known, the quantization condition for the WKB solution is obtained by matching boundary conditions between classically allowed and forbidden regions and leads to the Onsager quantization condition, which, in the absence of Berry phase, reduces to:

\be
\sum_s \frac{s}{eB}\int_{p_y^{-}}^{p_y^{+}} dp_y p^{\rm cl}_{x,s}(p_y,\epsilon)= l^2 	A(\epsilon_n)= 2\pi (n+1/2).
\ee

\noindent where we made the change of variables $p_y=-eB x$, $A(\epsilon_n)$ is the area in momentum space inside the iso-energetic contour $H(k_x,k_y)=\epsilon_n$ and $n$ is the Landau level index. From Eq.~\eqref{Pol} we write the Dipole as $D_y=-e l^{2} \int d x \ {\rm Im}(u^*_0(x) \partial_x u_0(x))$. There are two types of contributions to the dipole moment, namely the sum of the contributions coming from each classical root separately and the contribution corresponding to the interference-like crossed terms between the two classical roots. The semiclassical approximation is expected to work when the oscillations of the wavefunction occur over a length much smaller over that in which the confining potential changes, therefore we restrict to states that have a high energy and hence a large Landau level index. For these states the crossed terms have rapid oscillatory factors of the form $\exp\left(i/\hbar \int^x dx' (p^{\rm cl}_{x,1}-p^{\rm cl}_{x,2})\right)$ and can be neglected. In addition in this case the integrals can be approximated to be taken over the classically allowed region, because the states with rapid oscillations in the classically allowed region correspondingly have fast decay in the classically forbidden region. Therefore the terms in which the oscillatory part cancels within the classically allowed region have the dominant contribution. As a consquence we obtain:

\begin{equation}
\begin{split}
&\int d x \ {\rm Im}(u^*_0(x) \partial_x u_0(x)) \approx\\
& -\sum_s \frac{s}{eB}\int_{p_y^{-}}^{p_y^{+}} dp_y \frac{1}{|v_{x,s}^{\rm cl}(p_y,\epsilon)|}p^{\rm cl}_{x,s}(p_y,\epsilon)\\
&=- \oint dt \ p_x(t,\epsilon).\\
\end{split}
\end{equation}
%
\noindent where in the last line we have established the connection to the classical cyclotron motion described in the previous section, in which $d{\bf r}(t)/dt={\bf v}=l^2 d{\bf k}(t)/dt \times \hat{z} $, from which we have that $v_x=l^2 dk_y/dt$. Additionally we must normalize the WKB solution. Using the same approximations employed to compute the dipole one obtains that the approximate normalization of the WKB solution is:

\be
 \int d x \ u^*(x) u(x) \approx \sum_s \frac{1}{eB}\int_{p_y^{-}}^{p_y^{+}} dp_y \frac{1}{|v_{x,s}^{\rm cl}(p_y,\epsilon)|}= \oint dt.
\ee

\noindent Thus the normalization is simply the total time that takes to complete the classical cyclotron orbit. A similar analysis can be followed to find the dipole along the $x$ direction $D_x$. Therefore within the approximations here outlined, the WKB quantized wave-function predicts that the dipole moment is simply given by time averaged position of the electron over the semiclassical orbit:

\be
{\bf D}\approx-|e| \langle  {\bf r}(t,\epsilon_n) \rangle_t \equiv -|e| \frac{\oint dt \ {\bf r}(t,\epsilon_n)}{\oint dt},
\ee

\noindent where the integral is perfomed over the semiclassical cyclotron orbit ${\bf r}(t)=l^2 {\bf k}(t) \times \hat{z}$, with ${\bf k}(t)$ tracing a constant energy contour $H({\bf k}(t))=\epsilon_n$ at a speed $v=|\partial \epsilon/\partial {\bf k} |$~\cite{Kittel}. This picture predicts intuitively and generically that the dipole is orthogonal to the direction of the distortion of the Fermi surface since the real space orbit is a rotated version of the Fermi surface.

\section{Quantum Hall Ferroelectrics at the surface of Topological crystalline insulators}\label{TCI}

There are various material platforms for multi-valley systems of the type-I, where anisotropic valleys are located at non-time-reversal-invariant momenta and each valley lacks twofold symmetry. One candidate platform are multi-valley systems on the (111) surface of bismuth \cite{Bi2001,Bi111}. Another interesting platform, on which we will focus in this section, is the (001) surface of topological crystalline insulators (TCI) SnTe, Sn$_{x}$Pb$_{1-x}$Se and Sn$_{x}$Pb$_{1-x}$Te \cite{LiuSnTe}. In both Bi(111) and in TCI's, surface state Landau levels have been observed by means of scanning tunneling microscopy~\cite{Yazdani,Madhavan}.


To study ferroelectricity in at the surface of TCI's, we consider a model with two valleys ($M=2$) of Dirac fermions located at opposite momenta, at filling $\nu=1$. This model describes the low-energy dispersion of the [001] surface of SnTe and PbSnTe topological crystalline insulators in the low temperature phase~\cite{Serbyn}. Here, the Dirac cones are generically ``tilted"~\cite{Dipole} and described by the following Hamiltonian:
\be
H_0=\pm v_x\sigma_x p_x+v_y\sigma_y p_x\pm\delta v_x p_x,
\ee

\begin{figure}
	\begin{center}
		\includegraphics[width=3.3in]{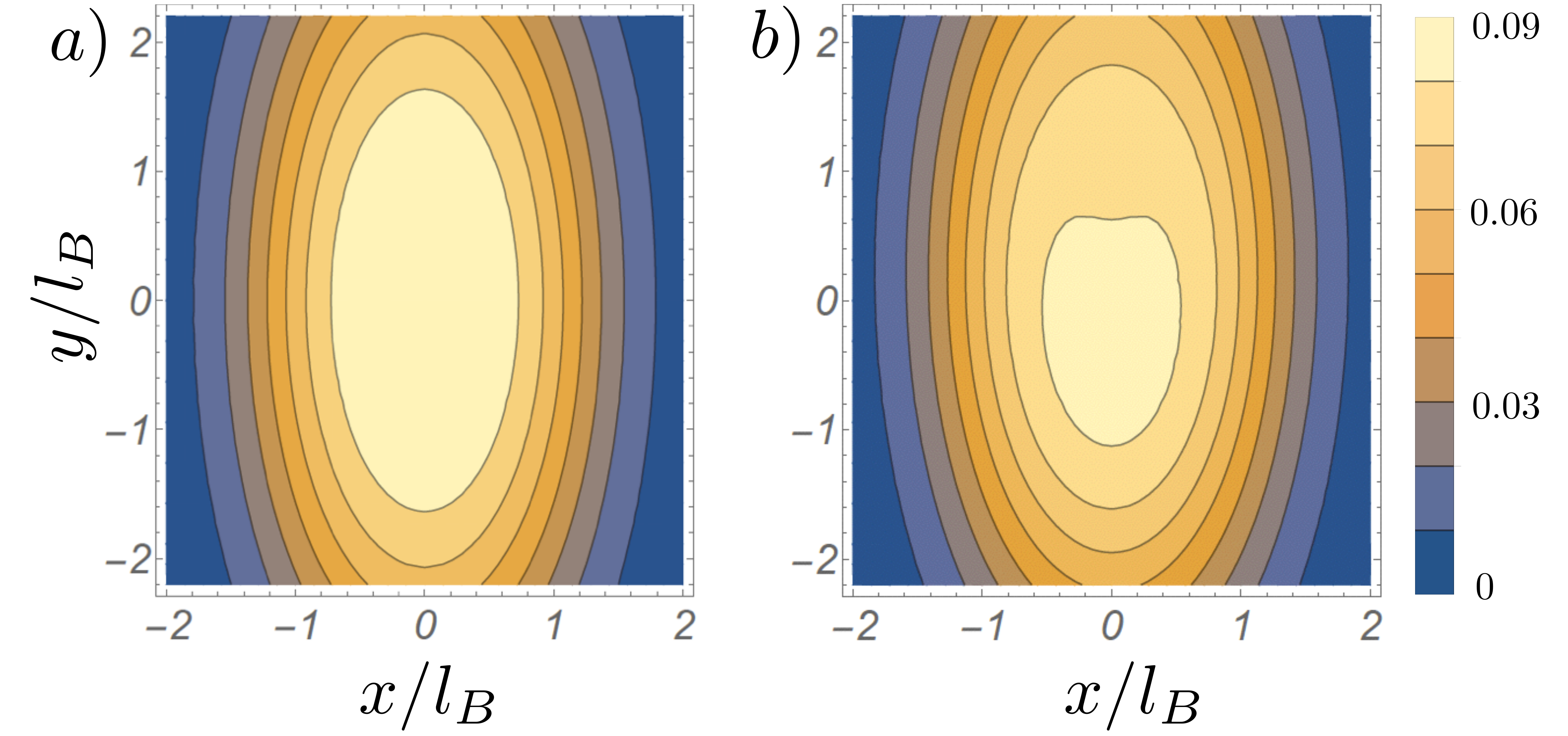}
	\end{center}
	\caption{(color online) Probability amplitude contours for coherent states in the first Landau level of a Dirac cone with anisotropic velocities $v_x/v_y\approx 2.2$ (a), and with anisotropy and tilt $\delta v_x \approx 0.3$ (b). The wavefunction carries a dipole moment perpendicular to the tilt of the Dirac cone.}
	\label{wavefs}
\end{figure}





\noindent The only discrete symmetry that leaves each valley invariant is a single mirror plane along the line that connects the Dirac cones~\cite{Serbyn,Dipole}. We consider the $n=1$ Landau level~\footnote{The zeroth Landau level does not carry dipole even in the presence of the tilt because the Hamiltonian retains an intra-valley particle-hole times inversion symmetry which forbids a dipole moment for the zeroth level. However, all other Landau levels will carry a dipole.}. The wavefunctions associated with each valley break rotation and inversion symmetry as illustrated in Fig.~\ref{wavefs}. Using the form factors described in Appendix~\ref{tilted} and applying Eq.~\eqref{xchange} to the case of $M=1$ at $\nu=1$, the Hartree-Fock energy can be found to be (up to a global constant):

\be
E/A=-\beta n_z^2,
\ee

\noindent where $n_z={\rm tr}(\sigma_z P)$ and $\sigma_z$ is a Pauli matrix in the valley indices, and $A$ is the system area. $\beta$ is a positive constant, in agreement with the theorem of full valley polarization, and it is explicitly given by:

\begin{equation}
\begin{split}
&\beta=\frac{\tau^2}{4 \pi}\int\frac{d^2 q}{(2\pi)^2} v_q e^{-|\bar{q}|^2/2} \left(3-\frac{ |\bar{q}|^2}{2}\right)^2   \bar{q}_y^{2},
\end{split}
\end{equation}


\noindent where $\tau=\delta v_x/(\sqrt{2} v_y)$, and $(\bar{q}_x,\bar{q}_y)=(\sqrt{v_y/v_x} q_y,\sqrt{v_x/v_y} q_x)$. Therefore, the ground state is an Ising type ($n_z=\pm1$) ferroelectric. The dipole moment, to leading order in the Dirac cone tilt, can be computed using the form of the Landau levels in terms of cyclotron raising and lowering operators (Appendix~\ref{tilted}), and using the Eq.~\eqref{Deq1}, and it is found to be:




\begin{equation}\label{DyDirac}
\begin{split}
&D_y\approx \pm  |e l | {\rm sign}(B)  \frac{3 \delta v_x}{\sqrt{2} v_y}, \ D_x=0.
\end{split}
\end{equation}

\noindent The dipole of each valley is orthogonal to the direction of the Dirac cone tilt and reverses with the direction of the perpendicular field. This is allowed since the magnetic field breaks the mirror symmetry present at zero field. In our discussion we have neglected the Zeeman coupling. This approximation is justified when the Zeeman energy is much smaller than the cyclotron spacing of the Dirac Landau levels in question, but including the Zeeman term should not change qualitatively the results in these systems.


\section{Experimental manifestations of quantum Hall ferroelectrics}

We now discuss potential experimental manifestations of the quantum Hall ferroelectrics. In the case of Bi(111) surface~\cite{Yazdani}, we expect these states to appear at odd integer fillings. The degree of broken inversion asymmetry in this system is expected to increase with Landau level index because the lowest Landau levels might be well described by an elliptical dispersion near the bottom of the hole bands, for which no dipole moment is expected, but only nematic states reported experimentally~\cite{Yazdani}. In this system, because the system has coexisting ferroelectric and nematic character, one signature is the appearance of additional ferroelectric domain walls in addition to those associated with nematicity seen at even fillings. One interesting possibility is the existence of gapless edge states at the ferroelectric domain walls. If one neglects inter-valley scattering, the valley pseudospin is conserved and one expects gapless charge carrying counter-propagating edge modes that arise at the domain walls. Due to their ferroelectric nature, these domains may be manipulated by STM bias voltage or in-plane external electric field. Finally, we note that Bi (111) hole pockets located at opposite momenta carry opposite in-plane spin-polarizations, which cancel when the two are equally occupied. Therefore, valley-polarized states at odd-integer filling also carry an in-plane spin polarization that could be manipulated with in-plane magnetic fields.  However, the broken inversion symmetry of this state would manifest directly in the charge distribution of quasiparticles which would carry a dipole moment given by the formulae described in section~\ref{moment}. We note in passing that other types of quantum Hall ferroelectrics arising in wide quantum wells~\cite{widewells} and in bilayer graphene~\cite{Barlas} have been previously proposed, but to our knowledge these proposals have not been experimentally realized so far.

\subsection{Measurement and physical meaning of the Dipole moment}\label{meas}

The dipole moment we have computed is defined as the change of the average position of every electron in a Landau level along a path that connects the Hamiltonian of interest with a reference initial Hamiltonian that has inversion symmetry. The argument presented in Section~\ref{expdipole} demonstrates that, as long as there is translational invariance, such a change is independent of the guiding center variables and hence it is the same for any orbital constructed within a given Landau level.

As we noted in section~\ref{moment}, the presence of the metallic edge of quantum Hall states prevents the build-up of charge at the boundaries of quantum Hall ferroelectrics. This may invalidate conventional approaches to measure the polarization used in ordinary ferroelectrics that rely on the measurement of macroscopic charge accumulation at the boundaries. Experimental manifestations of the formation of this inversion symmetry breaking quantum Hall state, therefore require looking at other kinds of observables. One instance of the manifestation of the broken inversion symmetry will be the charge distribution of quasiparticles. One consequence of inversion symmetry is that the charge distribution of elementary quasiparticles will be inversion symmetric. However, this will no longer be true when the system breaks inversion symmetry. Consider, the elementary quasi-hole of the system which is created by removing the electron from a single orbital $\psi_{k,J}({\bf r})$, with an intra-Landau label $k$, in the Slater determinant associated with completely filling a Landau level that we label by $J$. The particle density in the quasi-hole state is:

\be
\rho_{qh}({\bf r})=\frac{1}{2\pi l^2}-|\psi_{k,J}({\bf r})|^2,
\ee

\noindent where $1/{2\pi l^2}$ is the background density of the $\nu=1$ state. Now, from this equation we can see that the change of the average position of the quasi-hole state computed along an adiabatic path that connects the Hamiltonian of interest with a reference inversion symmetric Hamiltonian will coincide with the change of position of the single particle state $\psi_{k,J}({\bf r})$ with an overall opposite sign. Therefore the change of the dipole moment of the quasi-hole is exactly minus the dipole moment that we have described in Section~\ref{moment}.

Broadly speaking, the quasi-hole can be considered as a limiting version of a hole inside the sample that separates the quantum Hall ferroelectric and vacuum. When the size of this hole is very large there are gapless excitations that appear accompanying the formation of the chiral edge that will screen the build up of charge. However, when the size of the hole is small or comparable to the magnetic length these excitations will acquire a finite size gap and we expect that dipole moments will build up in such limit. Therefore, a dipole moment should build up for cavities or holes inside the sample that are on the order of the magnetic length, in resemblance to the {\it cavity electric fields} that are used as a classic method in elementary discussions of dielectrics as a way to measure the electric dipole moment~\cite{Zangwill}. Strictly speaking a cavity is different from a free quasi-hole because there is an explicit potential that expels electrons from its interior and breaks translational symmetry, which was one of our assumptions on the derivation of the dipole moment. Studying in detail the dipole moment build up in such cavities is an interesting open question that we hope future work will address.

\section{Quantum Hall nematics}\label{nemat}


In this section we consider multi-valley systems of type-II that develop nematicity but which carry no dipole moment. We focus on the case with two valleys whose anisotropic mass tensors are rotated by a $\pi/2$ angle relative to each other, as realized in AlAs~\cite{AlAsrev,Abanin}.  Our main objective is to study non-trivial charged excitations, but we briefly review ground state properties. At filling $\nu=1$ one finds that the energy is (up to a global constant)~\cite{Abanin}:

\be
E/A=-\alpha n_z^2.
\ee

\noindent where $n_z={\rm tr}(\sigma_z P)$, and $\alpha>0$. The system has Ising character in agreement with the theorem of full valley polarization. To further asses the validity of the Hartree-Fock approach we have performed DMRG simulations that are described in Appendix~\ref{dmrg}. It is in fact easy to covince one-self that the Ising nematic state is an exact eigenstate of the Hamiltonian projected into the valley degenerate Landau level. We confirmed that it is the ground state via DMRG employing a torus with a square aspect ratio (Appendix~\ref{dmrg}) for as many as $24$ electrons and for mass anisotropies as large as $m_x/m_y=16$.






\begin{figure}
	\begin{center}
		\includegraphics[width=3.3in]{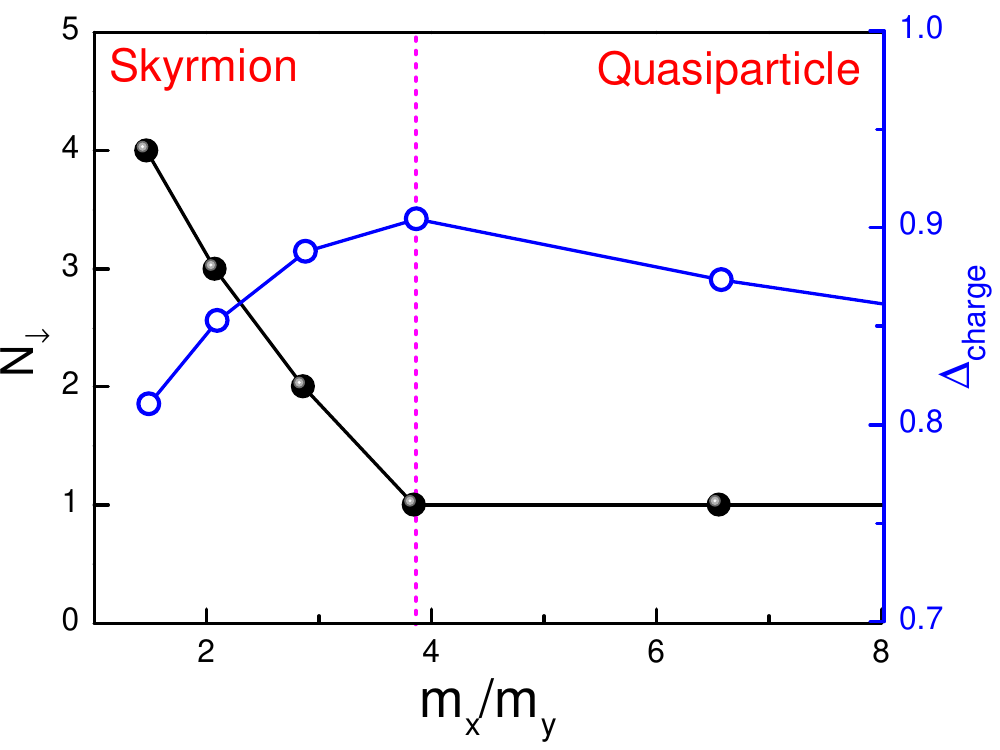}
	\end{center}
	\caption{(color online) Number of minority valley electrons (left axis) for quasiparticles as a function mass anisotropy in a two valley system (AlAs) and their charge gap (right axis). The charge gap is in Coulomb energy units of $e^2/(\epsilon l)$.}
	\label{gaps}
\end{figure}

\subsection{Valley skyrmions in quantum Hall nematics}\label{skyrm}

Having established that the ground state is a nematic state, we now turn to study its charged excitations. In the SU(2) invariant limit, {\it i.e.} when the mass tensors are identical for both valleys, we expect the lowest energy excitation to be infinite sized skyrmions~\cite{Sondhi}. When the mass tensors are slightly different, we expect that the skyrmions will have a finite size dictated by the competition between the Ising anisotropy which wants to shrink them and the Coulomb energy which wants to expand them to smear the charge over large distances~\footnote{The stiffness cost is scale invariant and hence indifferent to the size of the skyrmion.}. To be able to study accurately the properties of skyrmions we resort to DMRG~\footnote{The energetics can be captured by a non-linear sigma model~\cite{Sondhi,Moon}, but using long-wavelength estimates for non-linear sigma model parameters can lead to a substantial underestimation the size of the skrymions. This is a known pitfall of this type of estimates for small skyrmions~\cite{Abolfath}.}. The skyrmion quasi-electron can be obtained as the ground state of the Hamiltonian when $N=N_\phi+1$~\footnote{Quasi-hole and quasi-electron excitations are related by particle hole symmetry.}. Figure~\ref{gaps} shows that non-trivial skyrmions (those involving at least one spin flip) survive up to large mass ratio $m_x/m_y\approx 3.8$. AlAs has mass anisotropy of about $m_x/m_y\approx5$~\cite{AlAsrev}. The experiment of Ref.~\cite{AlAsskyrm} found a non-linear dependence of the charge gap on strain, which was interpreted as evidence for skyrmions. Our findings suggest that this is scenario is not unlikely since the critical mass ratio to observe skyrmions is ultimately dependent on the details of interactions and can easily change by effects beyond our model (e.g. finite well widths and Landau level mixing).

We have also computed the charge gaps for these quasiparticles in Fig.~\ref{gaps} . The charge gap of a system can be defined as the jump in chemical potentials. The chemical potential is the energy change for adding a particle. Thus the chemical potential to add a quasi-electron on top of the ground state is $\mu_+=E_{N=N_\phi+1}-E_{N=N_\phi}$, similarly the chemical potential for quasiholes is $\mu_-=E_{N=N_\phi}-E_{N=N_\phi-1}$. With this the charge gap is found to be:

\be
\Delta_{charge}=\mu_+-\mu_-=E_{N=N_\phi+1}+E_{N=N_\phi-1}-2 E_{N=N_\phi}.
\ee

\noindent A detailed comparison of the charge gaps of the non-trivial skyrmions we have found with those expected within Hartree-Fock theory is presented in Appendix~\ref{compa}. Finally, although we focused on the skyrmions on the nematic case, skyrmions could also be present in the ferroelectric states. These Skyrmions will carry also a dipole moment although its magnitude cannot be inferred from the simple formulae we have developed. The precise pattern of the skyrmion texture in this case might be very complex due to the interplay of breaking of inversion symmetry and the long range nature of Coulomb interactions. We hope future work addresses the nature of these interesting quasiparticles.




\section{Summary}

In summary, we have shown that multivalley systems with anisotropic dispersions lead generically to ferroelectric or nematic quantum Hall states at odd integer Landau level fillings. The ferroelectric states arise when the parent Fermi surface of a single valley lacks an inversion center, such as the case of Bi$(111)$~\cite{Yazdani}. We have shown that the resulting dipole moment has an intimate relation with the underlying fermi surface geometry of the parent metal. We also demonstrated the existence of non-trivial charged skyrmion excitations in the nematic states with an unexpectedly large stability to the Ising symmetry breaking terms, shedding light into the question of the presence of these excitations in AlAs~\cite{AlAsskyrm}.

\begin{acknowledgments}
We would like to thank Vidya Madhavan, Benjamin Feldman, Malika Randeria,  Ali Yazdani, Donna Sheng, Heun Mo Yoo, Joonho Jang, Ray Ashoori, Fengcheng Wu, and Allan H. MacDonald for stimulating discussions. We specially thank Pok Man Tam for discussions and for catching a mistake in the numerical prefactor of Eqs.~\eqref{DyDirac} and~\eqref{LL1} in an earlier unpublished version of the manuscript. While at MIT I.S. was supported by the Pappalardo Fellowship. Z.Z. is supported by the David and Lucile Packard foundation. L.F. is supported by the DOE Office of Basic Energy Sciences, Division of Materials Sciences and Engineering under Award No. DE-SC0010526.
\end{acknowledgments}

\appendix

\section{Landau levels and density form factors for anisotropic parabolic dispersions}\label{parabolic}
Consider the problem of Landau levels (LL) with an anisotropic mass tensor:
\be\label{Hpar}
H=\frac{1}{2m^*}p_a g_{ab} p_b.
\ee

\noindent Where $p=\nabla/i-e A$, and $g$ is a tensor that can be diagonalized as $g=Q^T S^2 Q$, where $Q\in SO(2)$ and $S$ is diagonal with positive eigenvalues and $\det S=1$. Explicitly $S={\rm diag}\{(m_x/m_y)^{1/4},(m_y/m_x)^{1/4}\}$ and $m^*=(m_x m_y)^{1/2}$. We can define rescaled momenta along the principal axes of the tensor to be: $\pi_a=(S Q)_{ab} p_b$, these satisfy the same commutation relations as the original ones:

\be
[\pi_a,\pi_b]=i l^{-2}\epsilon_{ab}.
\ee

\noindent Which allows to solve the LL problem by defining the LL raising operators:

\be\label{a}
a\equiv\frac{l}{\sqrt{2}}(\pi_x+i \pi_y),
\ee

\noindent satisfying $[a,a^\dagger]=1$. The guiding center operators are intra-Landau level operators defined as:

\be\label{R}
R_a\equiv r_a+l^2 \epsilon_{ab} p_b.
\ee

\noindent They satisfy:

\begin{equation}\label{Rcomm}
\begin{split}
& [R_a,R_b]=-i l^2 \epsilon_{ab},\\
& [R_a,p_b]=[R_a,\pi_b]=0.\\
\end{split}
\end{equation}

The single particle Hilbert space can be decomposed into a tensor product $|n\rangle\otimes |m\rangle$, where $|n\rangle$ are the Landau level indices on which ${a,a^\dagger}$ act and $|m\rangle$ are intra-Landau level indices on which $R_a$ acts. Now the projected interaction into the Landau level of interest is obtained by imagining we have two flavors of particles with different mass tensors ($g_1=Q_1^T S_1^2 Q_1$ and $g_2=Q_2^T S_2^2 Q_2$) which interact via a potential that depends only on interparticle distance:

\be
V(r_1-r_2)=\frac{1}{A} \sum_q V_q e^{i q\cdot(r_1-r_2)}.
\ee

\noindent where $r_1$ and $r_2$ above are understood to be operators. Using Eq.~\eqref{R} we decompose the position of each particle as:

\be
r_1 \equiv R_1-l^2 \epsilon p_1=R_1-l^2 \epsilon Q_1^T S_1^{-1} \pi_1,
\ee

\noindent where we are using matrix notation for the two component levi-civita symbol $\epsilon$. We have a similar expression for $r_2$. Using this we get:

\begin{equation}\label{densit}
\begin{split}
& e^{i q\cdot(r_1-r_2)}= e^{i q\cdot(R_1-R_2)} e^{-i l^2 (q_1 \cdot \pi_1-q_2 \cdot  \pi_2)},\\
& q_i=-S_i^{-1} Q_i \epsilon q.\\
\end{split}
\end{equation}

\noindent In the above expression the terms containing the operator $\pi_1$ produce inter-Landau level mixing and we proceed by projecting them into the zeroth landau level which is defined as $a |0\rangle=0$. By using Eq.~\eqref{a} in combination with the BCH formula one can show that:

\be\label{order}
e^{i l^2 q\cdot \pi}=e^{-\frac{l^2 |q|^2}{4}} e^{\frac{i l}{\sqrt{2}}\mathbf{q} a^\dagger} e^{\frac{i l}{\sqrt{2}}\mathbf{q^*} a},
\ee

\noindent where $\mathbf{q}=q_x+i q_y$. Then one obtains that the projected Hamiltonian is:

\be\label{P0vsupp}
P_0V(r_1-r_2) P_0=\frac{1}{A} \sum_q V_q e^{-l^2 \frac{ | q_1|^2+| q_2|^2}{4}} e^{i q\cdot(R_1-R_2)}.
\ee

\noindent Notice that $q_{1,2}$ are linear functions of $q$ described in Eq.~\eqref{densit}.

For higher Landau levels we need to modify the density form factors. Using Eq.~\eqref{order} and the algebra of raising and lowering operators one can show the following identities:

\begin{equation}\label{densities}
\begin{split}
& \langle n | e^{i l^2 q\cdot \pi} | n \rangle =\\
& = e^{-\frac{l^2 |q|^2}{4}} \sum_{m=0}^n \left(-\frac{l^2 |q|^2}{2}\right)^m \frac{n(n-1)...(n-m+1)}{(m!)^2},\\
& = e^{-\frac{l^2 |q|^2}{4}} L_n \left(\frac{l^2 |q|^2}{2}\right),
\end{split}
\end{equation}

\noindent where $n$ is the Landau level of interest and $L_n$ are the Laguerre polynomials. Therefore the interaction projected to the $n$ Landau level is:

\begin{equation}\label{densities}
\begin{split}
& P_nv(r_1-r_2) P_n=\frac{1}{A} \sum_q v_q L_n \left(\frac{l^2 |q_1|^2}{2}\right) L_n \left(\frac{l^2 |q_2|^2}{2}\right)\\
&  \times e^{-l^2 \frac{ | q_1|^2+| q_2|^2}{4}} e^{i q\cdot(R_1-R_2)}.\\
\end{split}
\end{equation}

\section{Landau levels and density form factors for anisotropic Dirac cones}\label{cones}

The derivation is very similar to that of Galilean electrons. We start from an anisotropic Dirac Hamiltonian:

\be\label{Hdirac}
H= \sigma_a g_{ab} p_b   =v \sigma'_a \pi_a .
\ee

\noindent where $g$ is a $3\times2$ tensor to which we apply a singular value decomposition of the form $g=R^T S Q$, where $R \in SO(3)$ describes a rotation of Pauli matrices $\sigma_a$ in pseudospin space, $Q \in SO(2)$ describes the transformation of the principal axes in real space, and $S$ is a $3\times2$ matrix whose upper $2\times2$ block is diagonal and characterizes the anisotropy of velocities $S={\rm diag}\{(v_x/v_y)^{1/2},(v_y/v_x)^{1/2}\}$, and whose lower row has zero entries, and $v=(v_xv_y)^{1/2}$. Using the same definition of ladder operators as in Eq.~\eqref{a}, the Hamiltonian and the spectrum are:

\begin{equation}\label{Hdirac2}
\begin{split}
& H= \frac{\sqrt{2} v}{l}\left(
\begin{array}{ccc}
 0 & a^\dagger \\
 a & 0 \\
\end{array}
\right), \\
& |n,k,s\rangle_D= \frac{1}{\sqrt{2}}\left(
\begin{array}{ccc}
 |n,k\rangle  \\
 s |n-1,k\rangle \\
\end{array}
\right), \ {\rm for} \ n > 0 \\
& |0,k\rangle_D=\left(
\begin{array}{ccc}
 |0,k\rangle   \\
 0 \\
\end{array}
\right), \ E_{nks}= s \frac{\sqrt{2} n v}{l}
\end{split}
\end{equation}

\noindent where $s=\pm$ labels negative and positive energy Landau levels and $k$ is the guiding center intra-Landau level index. From these expressions we can obtain the form factors of the interaction projected to the Dirac Landau levels. One can show that the zero Landau level is identical to the Galilean case, and for the excited Landau levels one gets the same expressions discussed in the previous section but with the modification:

\be
L_n \left(\frac{l^2 |q|^2}{2}\right) \rightarrow \frac{1}{2} \left(L_n \left(\frac{l^2 |q|^2}{2}\right)+L_{n-1} \left(\frac{l^2 |q|^2}{2}\right)\right)
\ee

\section{Landau levels and density form factors in tilted Dirac cones}\label{tilted}

To the Hamiltonian, $H$, appearing in Eqs.~\eqref{Hdirac},~\eqref{Hdirac2} we add a perturbation describing the tilt of the Dirac cone along the x-direction:

\be
H_1=\delta v_x \pi_x=\delta v_x\sqrt{\frac{v_x}{v_y}} \frac{(a+a^\dagger)}{\sqrt{2} l},
\ee

\noindent where $a$ is defined in Eq.~\eqref{a}. The (un-normalized) perturbed zero energy Landau level to first order in $\delta v_x$ is found to be:

\be
|0\rangle_1 = \left(
\begin{array}{ccc}
 |0\rangle  \\
 - \tau |0\rangle  \\
\end{array}
\right),
\ee

\noindent where $\tau=\delta v_x/(\sqrt{2} v_y)$. This verifies that the zero Landau level does not carry dipole as expected since particle-hole times inversion is a symmetry. Generically the $n$-th Landau level will break inversion symmetry however. The valence and conduction first Landau levels have a perturbed form:

\be\label{LL1}
|\pm 1\rangle_1 = \frac{1}{\sqrt{2}}\left(
\begin{array}{ccc}
 |1\rangle \pm \tau( |0\rangle -2 \sqrt{2}|2\rangle)   \\
 \pm |0\rangle -3 \tau|1\rangle  \\
\end{array}
\right).
\ee

As described in the main text we imagine now two Dirac valleys at opposite momenta and with tilts $\delta v_x$ of opposite sign and equal magnitude. Let us assume we are filling only one of the two first Landau levels. The form factor obtained from the above perturbative expression is:

\begin{equation}
\begin{split}
& F_\tau(q)\equiv  {}_1\langle 1 | e^{i l^2 q\cdot \pi'} | 1 \rangle_1 \approx \frac{1}{2}\biggl[f_{11}(q)+f_{00}(q)\\
& - 2 \tau (f_{10}(q)+f_{01}(q))- 2 \sqrt{2}\tau (f_{21}(q)+f_{12}(q))\biggr].
\end{split}
\end{equation}


\noindent where $f_{nm}(q)\equiv e^{-l^2|q|^2/4} \langle n | e^{i l^2 q\cdot \pi'} | m \rangle$. To leading order in $\tau$ we can write the form factor as:

\begin{equation}
\begin{split}
& F_\tau(q) \approx  e^{-l^2|q|^2/4} \left[(1-\frac{q^2}{4})- \sqrt{2}i \tau q_y (3-\frac{q^2}{2})\right].
\end{split}
\end{equation}

One valley will have a form factor corresponding to $\tau$ and the other corresponding to $-\tau$ since the tilts have opposite signs. We label them by $\alpha=\{+,-\}$. If one performs the calculation of the exchange energy with these form factors then one finds that the exchange integrals in Eq.~\eqref{xchange}:

\begin{equation}
\begin{split}
& X_{\alpha\beta}=\frac{1}{2}\int\frac{d^2 q}{(2\pi)^2} v_q F_{\alpha}(\bar{q}) F_{\beta}(\bar{q}), \ \bar{q}=-S^{-1} Q \epsilon q.
\end{split}
\end{equation}

The exchange energy can then be shown to be:

\begin{equation}
\begin{split}
E/N_\phi=\frac{(1-n_z^2)}{2} X_\tau+{\rm const}.
\end{split}
\end{equation}

\noindent with:

\begin{equation}
\begin{split}
X_\tau=&\frac{X_{++}+X_{--}-X_{+-}-X_{-+}}{2}\\
=&\tau^2\int\frac{d^2 q}{(2\pi)^2} v_q e^{-|\bar{q}|^2/2} \left(3-\frac{ |\bar{q}|^2}{2}\right)^2   \bar{q}_y^{2}
\end{split}
\end{equation}

\noindent Clearly for a repulsive interaction with $v_q>0$ we get that $X_\tau$ is strictly positive. Therefore the ground state is the Ising nematic ferroelectric state: $n_z=\pm1$.

\section{DMRG implementation}\label{dmrg}
Numerically we consider $N$ electrons moving along the surface of torus with a magnetic field perpendicular to its surface. $L_x$ ($L_y$) represents the circumference of the torus along $x$ ($y$) direction and they satisfy the relation $L_xL_y=2\pi N_\phi$, where $N_\phi$ represents the number of orbitals. Here, we set the magnetic length $l_B\equiv\sqrt{\hbar c/eB}$ as the unit of length. We choose a square torus with aspect ratio 1, i.e., $L_x=L_y$. Periodic boundary condition requires $k_y\equiv  {2\pi j}/L_y $ with $j=0,1,...,N_\phi-1$. The  Coulomb interaction on finite size system has the form
 \begin{equation}
V\left( {\bf{r}} \right) = \frac{1}{{{L_x}{L_y}}}\sum\limits_{\bf{q}} V(q) \exp \left( {i{\bf{qr}}} \right)
 \end{equation}
 with $V(q)= 2\pi {e^2}/{L_xL_y\varepsilon q}$, and the wave-vectors are chosen to guarantee the periodicity of the interaction $V\left( x+L_x,y \right)=V\left( x,y+L_y\right)=V(x,y)$.
The projected Coulomb interaction between particles $i,j$ into the Lowest Landau level of a valley with mass anisotropic dispersion has the form:
 \begin{equation}
 V\left( {{{\bf{R}}_i} - {{\bf{R}}_j}} \right) ={\sum\limits_{\bf{q}} {V\left( q \right)} } {e^{ - {q_1^2}/2}}{e^{i{\bf{q}}\left( {{{\bf{R}}_i} - {{\bf{R}}_j}} \right)}} ,
   \end{equation}

\noindent where $q_{1}$ is a linear functions of $q$ controlled by the anisotropic mass tensors described in Eq.~\eqref{densit}. The matrix elements take the explicit form,

\begin{equation}\label{matrix}
\begin{split}
&{V_{{j_1}{j_2}{j_3}{j_4}}} = {{\delta '}_{{j_1} + {j_2},{j_3} + {j_4}}}\frac{1}{{2{L_x}{L_y}}} \sum\limits_{{\bf{q}},{\bf{q}} \ne 0} {{{\delta '}_{{j_1} - {j_4},{{{q_y}{L_y}} \mathord{\left/
 {\vphantom {{{q_y}{L_y}} {2\pi }}} \right.
 \kern-\nulldelimiterspace} {2\pi }}}}} \times\\
& \frac{{2\pi {e^2}}}{{\varepsilon q}}\exp \left[ {{{ - {q_1^2}} \mathord{\left/
 {\vphantom {{ - {q^2}} 2}} \right.
 \kern-\nulldelimiterspace} 2} - i\left( {{j_1} - {j_3}} \right){{{q_x}{L_x}} \mathord{\left/
 {\vphantom {{{q_x}{L_x}} {{N_\phi }}}} \right.
 \kern-\nulldelimiterspace} {{N_\phi }}}} \right].
\end{split}
\end{equation}


\noindent Here, the Kronecker delta with the prime means that the equation is defined modulo $N_\phi$.  We also consider a uniform and positive background charge so that the Coulomb interaction at $q=0$ is absent~\cite{Yoshioka84}.

\noindent For the case of two anisotropic valleys, the Coulomb interaction includes both the intra-component and inter-component interactions. After projecting onto the lowest Landau level, the effective Hamiltonian reads as:

\begin{equation}\label{matrix2}
\begin{split}
V= & \sum\limits_{i < j} {\sum\limits_{\bf{q}} {V\left( q \right)} } {e^{ - {q_1^2}/2}}{e^{i{\bf{q}}\left( {{{\bf{R}}_{1,i}} - {{\bf{R}}_{1,j}}} \right)}}  \\
&+  \sum\limits_{i < j} {\sum\limits_{\bf{q}} {V\left( q \right)} } {e^{ - {q_2^2}/2}}{e^{i{\bf{q}}\left( {{{\bf{R}}_{2,i}} - {{\bf{R}}_{2,j}}} \right)}}\\
&+ \sum\limits_{i < j} {\sum\limits_{\bf{q}} {V\left( q \right)} }  {e^{ - (q_1^2+q_2^2)/4}}{e^{i{\bf{q}}\left( {{{\bf{R}}_{1,i}} - {{\bf{R}}_{2,j}}} \right)}}.
\end{split}
\end{equation}

\noindent Here, the first two terms are intra-component Coulomb interaction, and the last term is inter-component Coulomb interaction, and $q_{1,2}$ are the linear functions of $q$ controlled by the anisotropic mass tensors of each valley in described in Eq.~\eqref{densit}. A similar expression to Eq.~\eqref{matrix} holds for the explicit matrix elements.

Simulations are realized by mapping the single particle orbitals into one dimensional lattice. The advantage of the density matrix renormalization group (DMRG) method, compared to exact diagonalization, is that it can achieve reliable results for larger system sizes. In this manuscript, we impose momentum-space DMRG method on torus, where each site in $x$ direction corresponds to different states labeled by the momentum in $y$ direction. The standard procedure is similar to real-space DMRG, but since the non-zero Coulomb interaction exists between the orbitals satisfying the momentum conservation $j_1+j_2=j_3+j_4$ mod $N_\phi$, only small number of momentum sectors in each DMRG block contributes to the ground state in the initial process. To ensure the convergence, one has to  save more momentum sectors with keeping  a small number of states in these sectors during the initial process, which may become important in the later DMRG sweep, while the remaining states are selected following the standard DMRG algorithm to reach the required truncation accuracy.  For our calculation, we keep around $2000\sim8000$ states and perform measurement during each DMRG sweep until the results converge.

\begin{figure}[htbp]
\begin{center}
\includegraphics[width=0.45\textwidth]{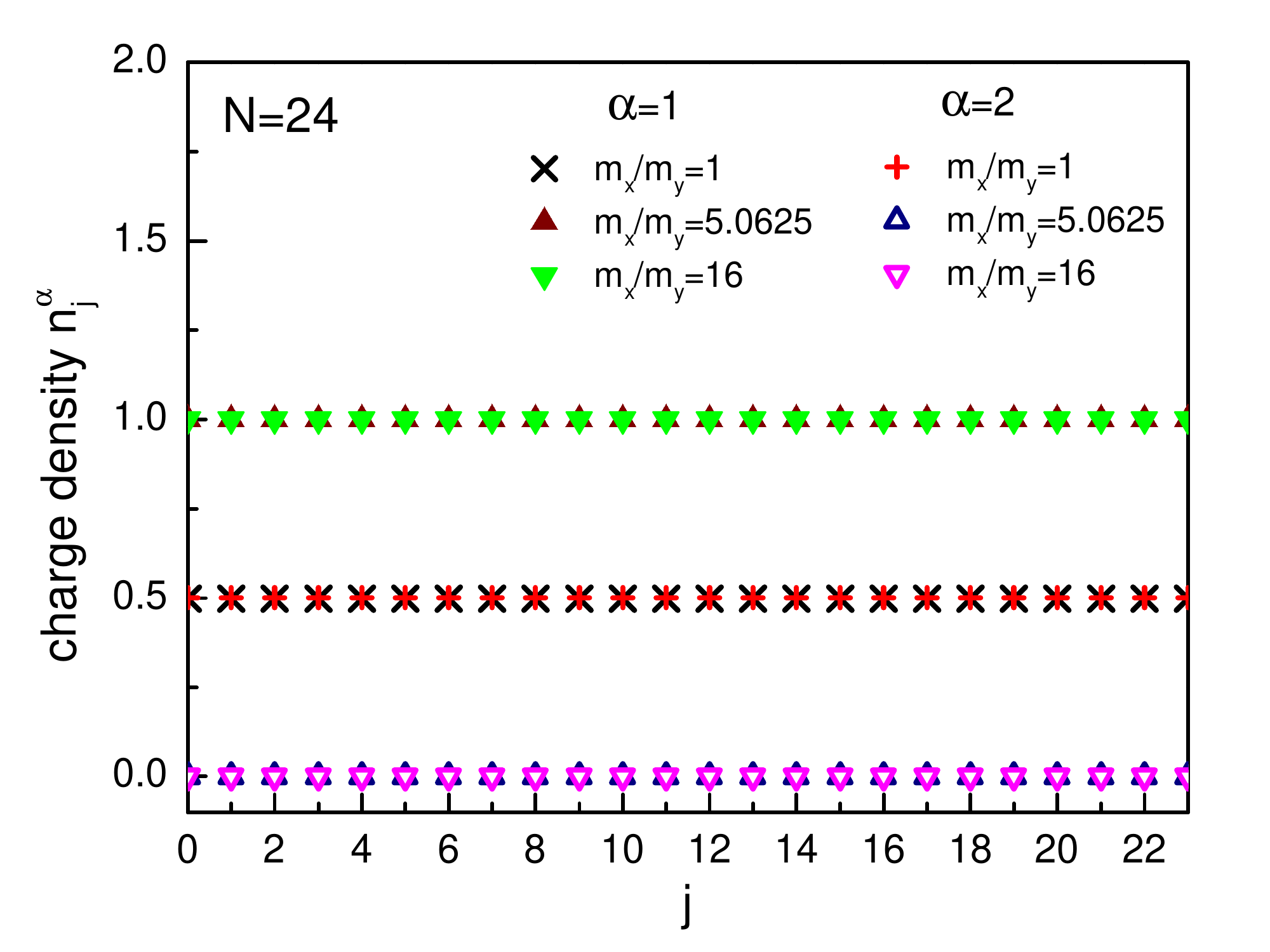}
\end{center}
\par
\renewcommand{\figurename}{Fig.}
\caption{ The charge density distribution for $N=24$ ($N_\phi=N$) system with anisotropic ratio $m_x/m_y=1, 5.0625, 16$. All of electrons tend to stay in the same flavor $\alpha$ orbitals ($j=0,2,\ldots, N_\phi-1$)  for anisotropic case.  $\alpha=1,2$ represent two components.}
\label{Fig:ChargeDensity_N24}
\end{figure}

We found that in the ground state the charge distributes uniformly into the two orbitals for isotropic case~\footnote{The method selects a single sector even in this case in which there are SU(2) related degenerate copies.}, i.e., $m_x/m_y=1$, while the ground state is realized in a sector in which all electrons polarize into a single orbital with the same flavor for anisotropic case, as shown in the Fig.~\ref{Fig:ChargeDensity_N24} for $N=N_\phi=24$ system with $(m_x/m_y)^{1/4}=1.5, 2$.

\begin{figure}[htbp]
\begin{center}
\includegraphics[width=0.45\textwidth]{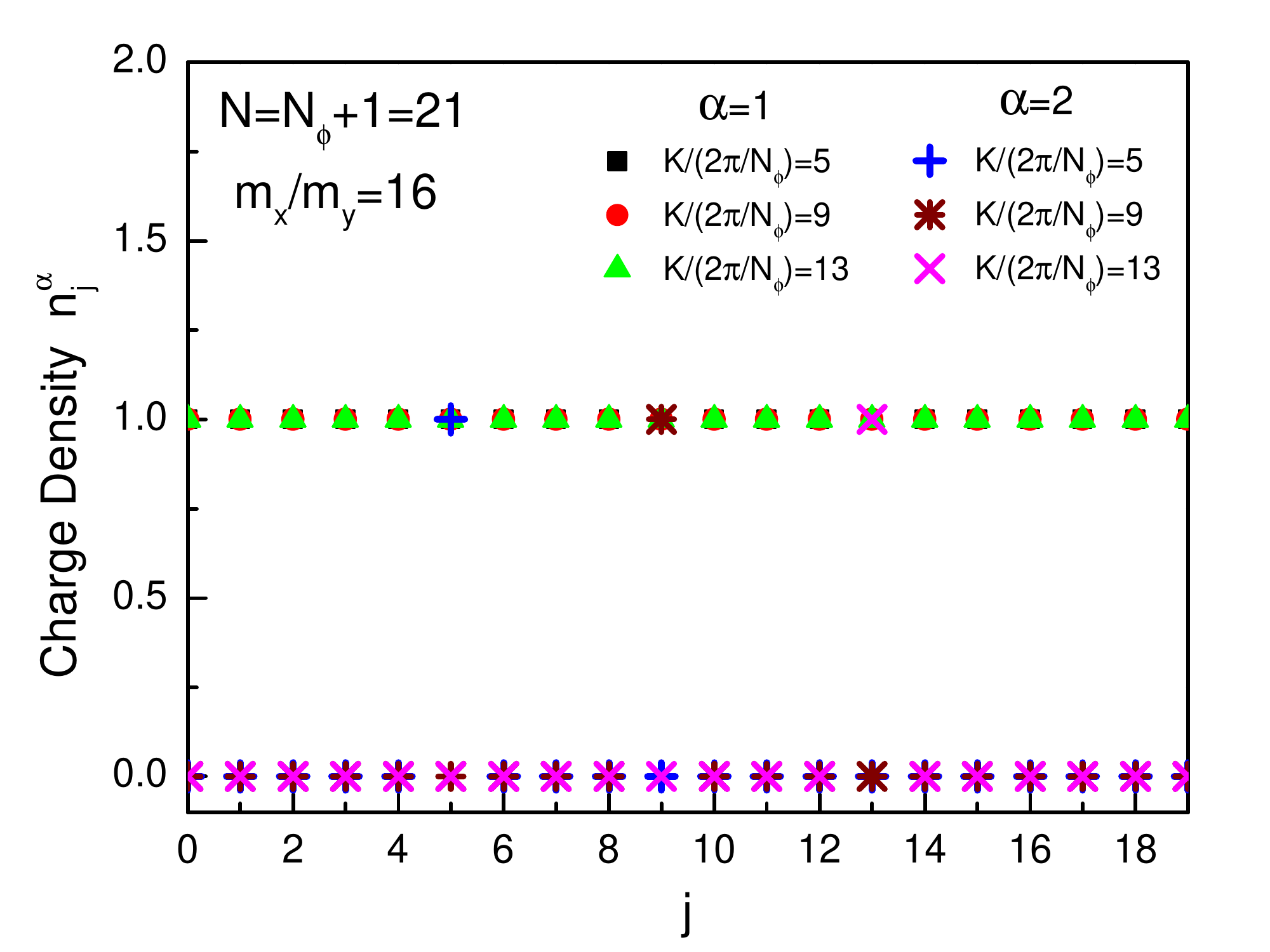}
\end{center}
\par
\renewcommand{\figurename}{Fig.}
\caption{ The charge density distribution for $N=21$ ($N_\phi=20$) system with anisotropic ratio $m_x/m_y=16$. The extra electron will stay in the orbital $K/(2\pi/N_\phi)$ if one target the corresponding total momentum, and the ground state energies are the same for different targeting momentum sectors.The total electron number for two flavors are $N_1=20$ and $N_2=1$.}
\label{Fig:ChargeDensity_N21r2}
\end{figure}
\begin{figure}[htbp]
\begin{center}
\includegraphics[width=0.45\textwidth]{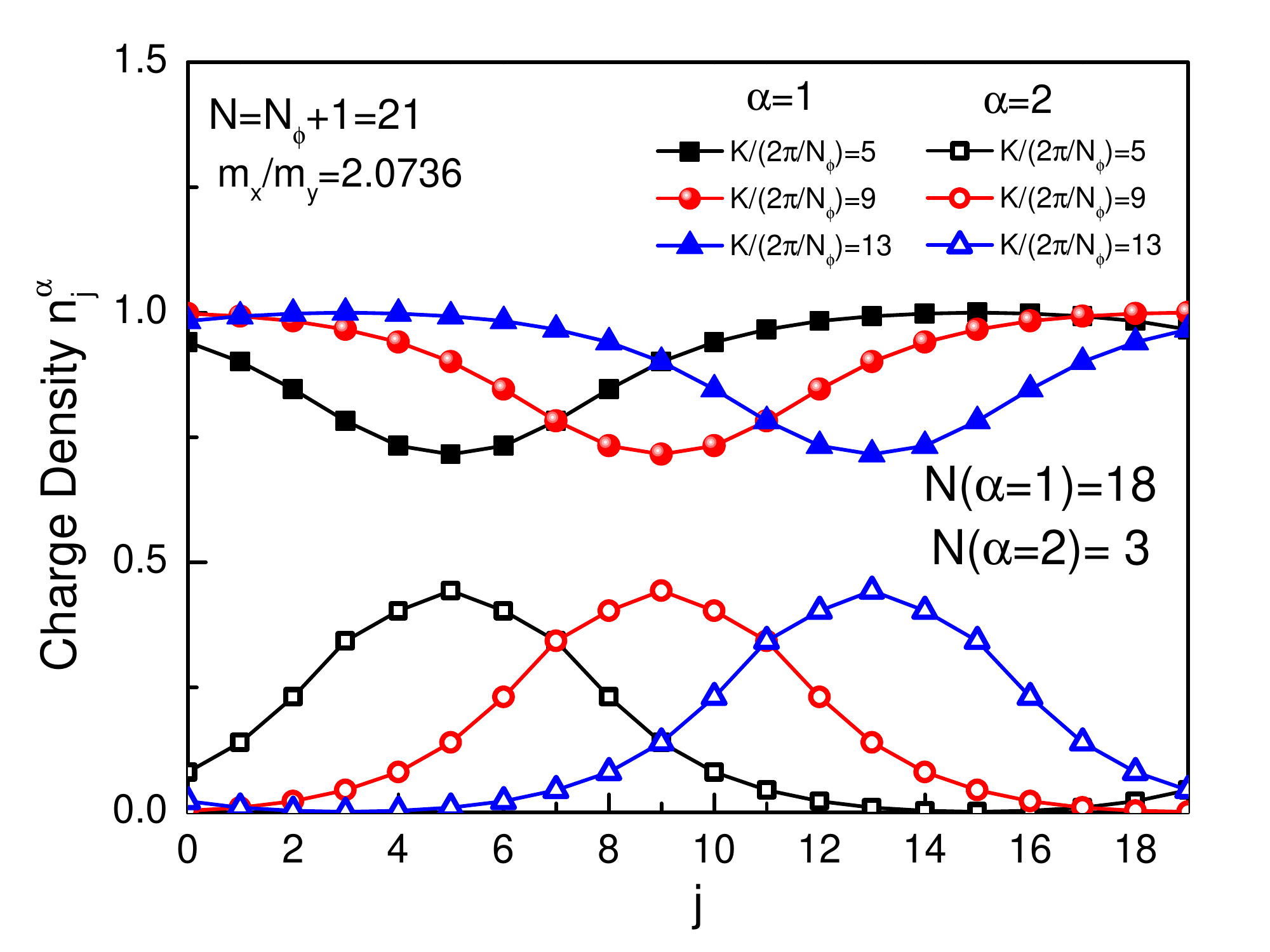}
\end{center}
\par
\renewcommand{\figurename}{Fig.}
\caption{ The charge density distribution for $N=21$ ($N_\phi=20$) system with anisotropic ratio $m_x/m_y=2.0736$. The extra electron will stay near the orbital $K/(2\pi/N_\phi)$ if one target the corresponding  total momentum, and the ground state energies are the same for different targeting momentum sectors. The electron number for two flavors are $N_1=18$ and $N_2=3$, which depends on the system size.}
\label{Fig:ChargeDensity_N21r1.2}
\end{figure}
\begin{figure}[htbp]
\begin{center}
\includegraphics[width=0.45\textwidth]{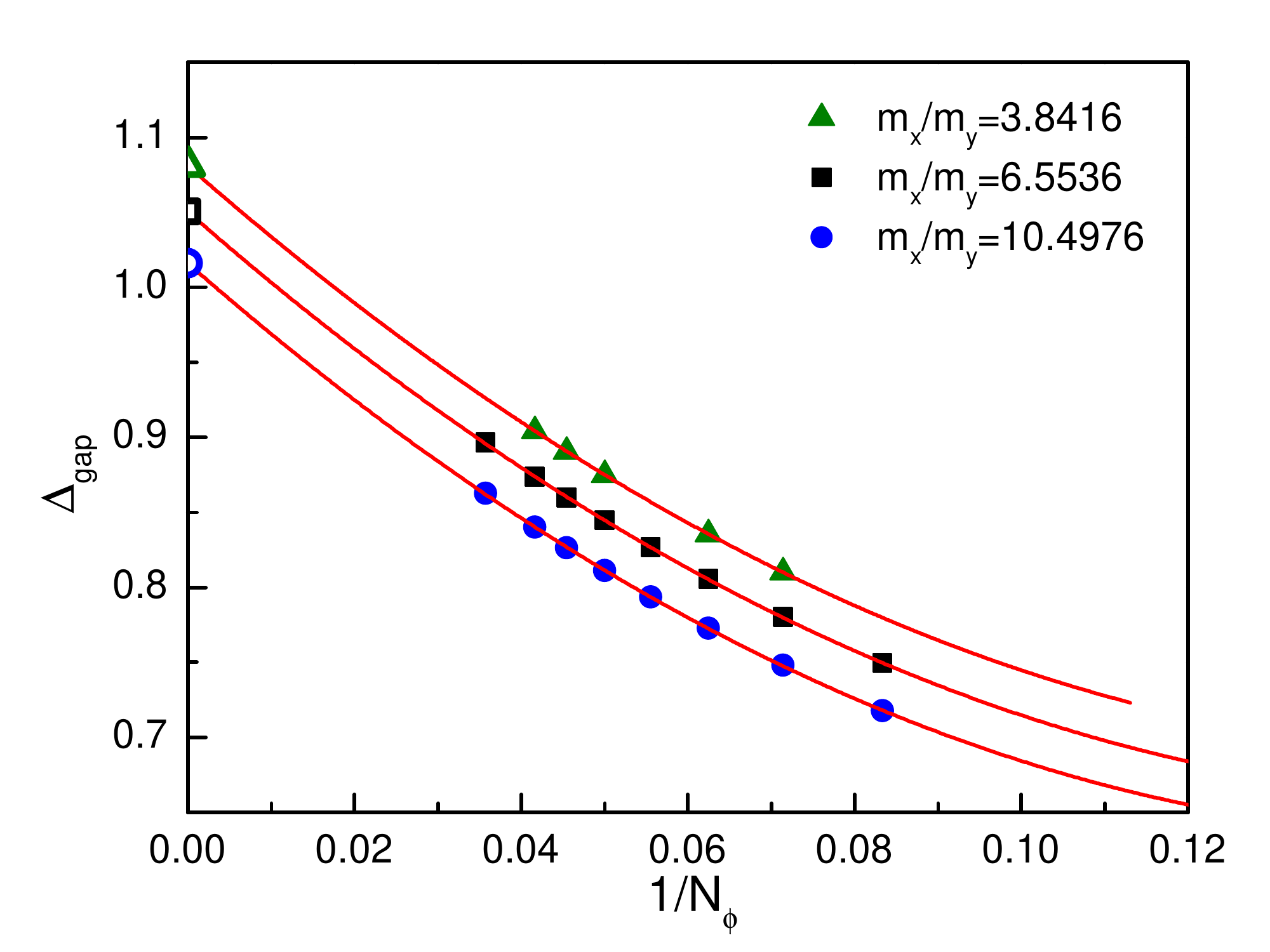}
\end{center}
\par
\renewcommand{\figurename}{Fig.}
\caption{ The finite size scaling of charge gap at quasiparticle regime for  $m_x/m_y=6.5536$ and $m_x/m_y=10.4976$. The system size ranges from $N_\phi=12$ to $N_\phi=28$. }
\label{Fig:ScalingGap}
\end{figure}

To study charged excitations we search for ground states with one particle added to those described above. The behavior of the charged excitations is markedly distinct for large and small anisotropic ratio $m_x/m_y$. When the anisotropic ratio is large enough, as shown in Fig.~\ref{Fig:ChargeDensity_N21r2} for $N=N_\phi+1=21$ system with $(m_x/m_y)^{1/4}=2$, $N_\phi$ electrons will fully occupy the orbitals with the same flavor, while the extra one will stay at the orbital with another flavor. The position of this extra electron is determined by the total momentum $K/(2\pi/N_\phi)$ targeted in DMRG simulation, but different momentum sectors have the same energy due to translational invariance. This is consistent with the behavior of the conventional quasiparticle predicted by Hartree-Fock analysis. However, for smaller $m_x/m_y$ ratio, as shown in Fig.~\ref{Fig:ChargeDensity_N21r1.2} with $(m_x/m_y)^{1/4}=1.2$, the total number of electrons in the majority flavor will be smaller than $N_\phi$, while he number of particles in the minority flavor will be larger than one, indicating that the quasiparticle has a nontrivial pseudo-spin texture.  Here, the number of particles in each flavor  depends on both anisotropic ratio $m_x/m_y$ and total number of particles $N$, and therefore extrapolations to $N\rightarrow \infty$ are required. Fig.~\ref{Fig:ScalingGap} shows the finite size scaling of the charge gap in the quasiparticle regime. We calculate the charge gap $\Delta_{gap}$ for the systems with $N_\phi=12$ to  $N_\phi=28$ and the anisotropic ratio  $(m_x/m_y)^{1/4}=1.6$ and  $(m_x/m_y)^{1/4}=1.8$. We fit the data as a quadratic function of $1/N_\phi$, and estimate the thermodynamic limit of the gap by extrapolating $1/N_\phi\rightarrow 0$.


\section{Comparison between Hartree-Fock and DMRG quasiparticle gap}\label{compa}

The charge gap in Hartree-Fock theory can be shown to coincide with minus twice the energy per particle of the ground state, which, for the Coulomb interaction, can be found to be:


\be
\Delta^{HF}_{\rm charge}= -2 \frac{E^0_{HF}}{N_\phi}=\frac{K(1-m_y/m_x)}{(m_x/m_y)^{1/4}} \sqrt{\frac{2}{\pi}} \frac{e^2}{l}.
\ee

\begin{figure}
	\begin{center}
		\includegraphics[width=3.2in]{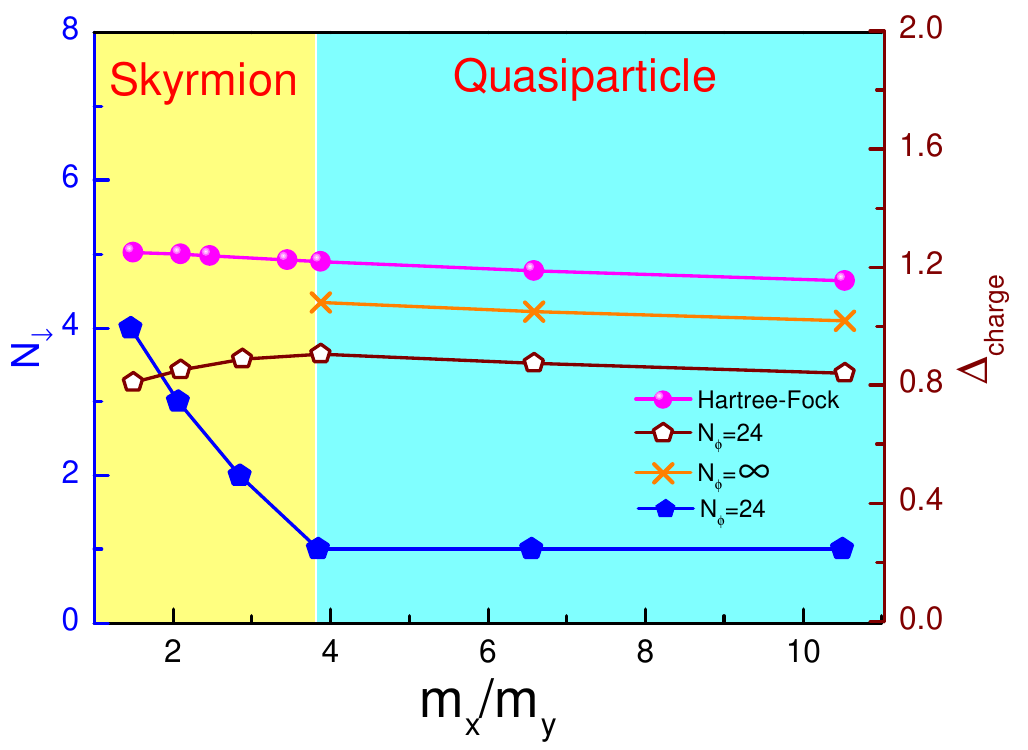}
	\end{center}
	\caption{The left vertical axis shows the number of spin flips (blue) involved in the lowest energy charged quasiparticle as a function of mass anisotropy for a system with two anisotropic pockets like AlAs. The right vertical axis shows different calculations of the charge gap including Hartree-Fock (purple) and DMRG for the largest system size (white) and the extrapolated values to infinite size (yellow).}
	\label{FigHF}
\end{figure}

\noindent Here $E^0_{HF}$ is the Hartree-Fock energy of the ground state with no quasiparticles and $K$ is the elliptic integral of the first kind. This formula corrects a typo in Ref.~\onlinecite{Abanin}. Figure~\ref{FigHF} illustrates the Hartree-Fock charge gap for the conventional quasi-electron and quasi-hole pair computed within Hartree-Fock theory and compared to the results from DMRG. After extrapolation to infinite sizes the results show good agreement in the large mass anisotropy regime where quasi-particles are no longer expected to be skyrmions.

The physical reason for the decreasing gap at large anisotropies is that as the anisotropy increases the overlap between orbitals becomes smaller and their exchange energy gain is reduced. Because the energy gain per particle is reduced it is also less energetically costly to add or remove particles, making the gap a decreasing function of mass anisotropy.

\begin{thebibliography}{longbibliography}

\bibitem{Yazdani} B. E. Feldman, M. T. Randeria, A. Gyenis, F. Wu, H. Ji, R. J. Cava, A. H. MacDonald, and A. Yazdani, \emph{Observation of a nematic quantum Hall liquid on the surface of bismuth}, Science {\bf 354}, 316 (2016).

\bibitem{AlAsrev} M. Shayegan, E. P. De Poortere, O. Gunawan, Y. P. Shkolnikov, E. Tutuc, K. Vakili, \emph{Two-dimensional electrons occupying multiple valleys in AlAs}, Phys. Stat. Sol. b {\bf 243}, 3629 (2006).

\bibitem{Novoselov1} K. S. Novoselov, A. K. Geim, S. V. Morozov, D. Jiang, M. I. Katsnelson, I. V. Grigorieva, S. V. Dubonos, and A. A. Firsov, \emph{Two-dimensional gas of massless Dirac fermions in graphene}, Nature {\bf 438}, 197 (2005).

\bibitem{Kim} Y. Zhang, Y-W. Tan, H. L. Stormer, and P. Kim, \emph{Experimental observation of the quantum Hall effect and Berry's phase in graphene}, Nature {\bf 438}, 201 (2005).

\bibitem{Novoselov} K. S. Novoselov, E. McCann, S. V. Morozov, V. I. Fal'ko, M. I. Katsnelson, U. Zeitler, D. Jiang, F. Schedin, and A. K. Geim, \emph{Unconventional quantum Hall effect and Berry¡¯s phase of 2$\pi$ in bilayer graphene}, Nature Physics {\bf 2}, 177 (2006).

\bibitem{Si111} K. Eng, R. N. McFarland, and B. E. Kane, \emph{Integer Quantum Hall Effect on a Six-Valley Hydrogen-Passivated Silicon (111) Surface}, Phys. Rev. Lett. {\bf 99}, 016801 (2007).

\bibitem{PbTe} V. A. Chitta, W. Desrat, D. K. Maude, B. A. Piot, N. F. Oliveira Jr., P. H. O. Rappl, A. Y. Ueta, E. Abramof, \emph{Integer quantum Hall effect in a PbTe quantum well}, Physica E: Low-dimensional Systems and Nanostructures {\bf 34}, 124 (2006).

 \bibitem{Madhavan} Yoshinori Okada, Maksym Serbyn, Hsin Lin, Daniel Walkup, Wenwen Zhou, Chetan Dhital, Madhab Neupane, Suyang Xu, Yung Jui Wang, R. Sankar, Fangcheng Chou, Arun Bansil, M. Zahid Hasan, Stephen D. Wilson, Liang Fu, Vidya Madhavan, \emph{Observation of Dirac Node Formation and Mass Acquisition in a Topological Crystalline Insulator}, Science 341, 6153 (2013).

\bibitem{Abanin} D. A. Abanin, S. A. Parameswaran, S. A. Kivelson, and S. L. Sondhi, \emph{Nematic valley ordering in quantum Hall systems}, Phys. Rev. B {\bf 82}, 035428 (2010); A. Kumar, S. A. Parameswaran, and S. L. Sondhi, \emph{Microscopic theory of a quantum Hall Ising nematic: Domain walls and disorder}, Phys. Rev. B {\bf 88}, 045133 (2013).

\bibitem{AllanSnTe} X. Li, F. Zhang, A. H. MacDonald, \emph{SU(3) Quantum Hall Ferromagnetism in SnTe}, Phys. Rev. Lett. {\bf 116}, 026803 (2016).

\bibitem{Bi2001} C. R. Ast and H. H\"ochst, \emph{Fermi Surface of Bi(111) Measured by Photoemission Spectroscopy}, Phys. Rev. Lett. {\bf 87}, 177602 (2001).

\bibitem{Bi111} Y. Ohtsubo, J. Mauchain, J. Faure, E. Papalazarou, M. Marsi, P. Le Fevre, F. Bertran, A. Taleb-Ibrahimi, and L. Perfetti, \emph{Giant Anisotropy of Spin-Orbit Splitting at the Bismuth Surface}, Phys. Rev. Lett. {\bf 109}, 226404 (2012).

\bibitem{PbSnSe} P. Dziawa, B. J. Kowalski, K. Dybko, R. Buczko, A. Szczerbakow, M. Szot, E. Lusakowska, T. Balasubramanian, B. M. Wojek, M. H. Berntsen, O. Tjernberg and T. Story, \emph{Topological crystalline insulator states in $Pb_{1-x}Sn_{x}Se$}, Nature Materials 11, 1023 (2012).

\bibitem{Sondhi} S. L. Sondhi, A. Karlhede, S. A. Kivelson, and E. H. Rezayi, \emph{Skyrmions and the crossover from the integer to fractional quantum Hall effect at small Zeeman energies}, Phys. Rev. B {\bf 47}, 16419 (1993).

\bibitem{AlAsskyrm} Y. P. Shkolnikov, S. Misra, N. C. Bishop, E. P. De Poortere, and M. Shayegan, \emph{Observation of Quantum Hall ``Valley Skyrmions"}, Phys. Rev. Lett. {\bf 95}, 066809 (2005).

\bibitem{Kharitonov} M. Kharitonov, \emph{Phase diagram for the $\nu=0$  quantum Hall state in monolayer graphene}, Phys. Rev. B \textbf{85}, 155439 (2012); M. Kharitonov, \emph{Edge excitations of the canted antiferromagnetic phase of the $\nu=0$ quantum Hall state in graphene: A simplified analysis}, Phys. Rev. B \textbf{86}, 075450 (2012).

\bibitem{Murthy} G. Murthy, E. Shimshoni, and H. Fertig, \emph{Spin-Valley Coherent Phases of the $\nu=0$ Quantum Hall State in Bilayer Graphene}, arXiv:1709.02447 (2017).



\bibitem{Moon} K. Moon, H. Mori, K. Yang, S. M. Girvin, A. H. MacDonald, L. Zheng, D. Yoshioka, and S-C. Zhang, \emph{Spontaneous interlayer coherence in double-layer quantum Hall systems: Charged vortices and Kosterlitz-Thouless phase transitions}, Phys. Rev. B {\bf 51}, 5138 (1995).

\bibitem{Shibata2001}N. Shibata and D. Yoshioka, \emph{Ground-State Phase Diagram of 2D Electrons in a High Landau Level: A Density-Matrix Renormalization Group Study}, Phys. Rev. Lett. \textbf{86}, 5755 (2001).

\bibitem{Feiguin2008}A. E. Feiguin, E. Rezayi, C. Nayak, and S. Das Sarma, \emph{Density Matrix Renormalization Group Study of Incompressible Fractional Quantum Hall States}, Phys. Rev. Lett. 100, 166803 (2008).

 \bibitem{Xiang1996}T. Xiang, Density-matrix renormalization-group method in momentum space, Phys. Rev. B \textbf{53}, R10445 (1996).

 \bibitem{White1992} S. R. White, \emph{Density matrix formulation for quantum renormalization groups}, Phys. Rev. Lett. \textbf{69}, 2863 (1992).

 \bibitem{Zhao2011}J. Zhao, D. N. Sheng, and F. D. M. Haldane, \emph{Fractional quantum Hall states at 1/3 and 5/2 filling: Density-matrix renormalization group calculations}, Phys. Rev. B \textbf{83}, 195135 (2011).


 \bibitem{LiuSnTe} Junwei Liu, Wenhui Duan, and Liang Fu, \emph{Two types of surface states in topological crystalline insulators}, Phys. Rev. B 88, 241303 (2013).



 \bibitem{Serbyn} M. Serbyn and L. Fu, \emph{Symmetry breaking and Landau quantization in topological crystalline insulators}, Phys. Rev. B {\bf 90}, 035402 (2014).

\bibitem{Dipole} I. Sodemann and L. Fu, \emph{Quantum Nonlinear Hall Effect Induced by Berry Curvature Dipole in Time-Reversal Invariant Materials}, Phys. Rev. Lett. {\bf 115}, 216806 (2015).

 \bibitem{Thouless} D. J. Thouless, \emph{Quantization of particle transport}, Phys. Rev. B {\bf 27}, 6083 (1983).

\bibitem{Vanderbilt} R. D. King-Smith and D. Vanderbilt, \emph{Theory of polarization of crystalline solids}, Phys. Rev. B {\bf 47}, 1651(R) (1993).

\bibitem{Falkovs} L. A. Falkovskii, \emph{Quasiclassical Quantization of Electrons and Holes in Bismuth in a Magnetic Field}, J. Exptl. Theoret. Phys. (U.S.S.R.) {\bf 49}, 609 (1965);  A. Yu. Ozerin and L. A. Falkovsky, \emph{Berry phase, semiclassical quantization, and Landau levels}, Phys. Rev. B {\bf 85}, 205143 (2012).

\bibitem{Kittel} C. Kittel, \emph{Introduction to solid state}, John Wiley \& Sons, (1966).

\bibitem{Onsager} L. Onsager, \emph{Interpretation of the de Haas-van Alphen effect}, Phil. Mag. {\bf 43}, 1006 (1952).


\bibitem{widewells} D-W. Wang, E. Demler, and S. Das Sarma, \emph{Spontaneous symmetry breaking and exotic quantum orders in integer quantum Hall systems under a tilted magnetic field}, Phys. Rev. B {\bf 68}, 165303 (2003).

\bibitem{Barlas} Y. Barlas, R. Cote, K. Nomura, and A. H. MacDonald, \emph{Intra-Landau-Level Cyclotron Resonance in Bilayer Graphene}, Phys. Rev. Lett. {\bf 101}, 097601 (2008).


\bibitem{Zangwill} A. Zangwill. \emph{Modern electrodynamics}. Cambridge University Press, 2013.

\bibitem{BoYang} B. Yang, Z. Papi\'c, E. H. Rezayi, R. N. Bhatt, and F. D. M. Haldane, \emph{Band mass anisotropy and the intrinsic metric of fractional quantum Hall systems}, Phys. Rev. B {\bf 85}, 165318 (2012).

\bibitem{Moon} K. Moon, H. Mori, Kun Yang, S. M. Girvin, A. H. MacDonald, L. Zheng, D. Yoshioka, and Shou-Cheng Zhang, \emph{Spontaneous interlayer coherence in double-layer quantum Hall systems: Charged vortices and Kosterlitz-Thouless phase transitions}, Phys. Rev. B \textbf{51}, 5138 (1995).



\bibitem{Abolfath} M. Abolfath, J. J. Palacios, H. A. Fertig, S. M. Girvin, and A. H. MacDonald, \emph{Critical comparison of classical field theory and microscopic wave functions for skyrmions in quantum Hall ferromagnets}, Phys. Rev. B {\bf 56}, 6795 (1997).

 \bibitem{Yoshioka84}Daijiro Yoshioka, \emph{Ground state of the two-dimensional charged particles in a strong magnetic field and the fractional quantum Hall effect}, Phys. Rev. B \textbf{29}, 6833(1984).






\end{thebibliography}

\end{document}